\documentclass[journal,draftcls,onecolumn,12pt,twoside]{IEEEtranTCOM}
\usepackage{graphicx}
\usepackage[font=footnotesize]{caption}
\usepackage{cite}
\usepackage[none]{hyphenat}
\usepackage{hyperref}
\usepackage[cmex10]{amsmath}
\usepackage{mdwmath}
\usepackage{amssymb}
\usepackage{dsfont,mathrsfs,amsfonts}
\usepackage{subfig}
\ifCLASSINFOpdf
  
\else
 
\fi

\begin{document}
	
	\title{\huge{Bidirectional User Throughput Maximization Based\\\vspace{-0.45cm} on Feedback Reduction in LiFi Networks}}

	\author{\normalsize{\IEEEauthorblockN{Mohammad Dehghani Soltani, %~\IEEEmembership{ \normalsize{Student Member,~IEEE,}}
			Xiping Wu, %~\IEEEmembership{\normalsize{Member,~IEEE,}}
			Majid Safari, %~\IEEEmembership{\normalsize{Member,~IEEE,}} 
			and
			Harald Haas %~\IEEEmembership{\normalsize{Member,~IEEE}}}
			}}
		\vspace{-.6cm}
		%\IEEEauthorblockA{\IEEEauthorrefmark{}Institute of Digital Communication, University of Edinburgh, Edinburgh, UK\\
		%Email: \{M.Dehghani, Xiping.Wu, Majid.Safari, H.Haas\}@ed.ac.uk}
		\thanks{%Manuscript received April 15, 2017. 
			This work has been submitted to IEEE Transactions on Communications.
			The authors are with the LiFi Research and Development Center, Institute for Digital Communications, The University of Edinburgh. (E-mail: m.dehghani@ed.ac.uk; 
			xiping.wu@ed.ac.uk; majid.safari@ed.ac.uk; h.haas@ed.ac.uk).}
}
	
	% The paper headers
%	\markboth{Journal of Transactions on Communications,~Vol.~99, No.~99, June~2017}
%	{Soltani \MakeLowercase{\textit{et al.}}: Bidirectional User Throughput Maximization Based on Feedback Reduction in LiFi Networks}
	
	\maketitle
\vspace{-0.9cm}	
\begin{abstract}
\vspace{-0.5cm}
Channel adaptive signalling, which is based on feedback, can result in almost any performance metric enhancement. 
Unlike the radio frequency (RF) channel, the optical wireless communications (OWCs) channel is fairly static. This feature enables a potential improvement of the bidirectional user throughput by reducing the amount of feedback. 
Light-Fidelity (LiFi) is a subset of OWCs, and it is a bidirectional, high-speed and fully networked wireless communication technology where visible light and infrared are used in downlink and uplink respectively.
In this paper, two techniques for reducing the amount of feedback in LiFi cellular networks are proposed, \textit{i}) Limited-content feedback (LCF) scheme based on reducing the content of feedback information and \textit{ii}) Limited-frequency feedback (LFF) based on the update interval scheme that lets the receiver to transmit feedback information after some data frames transmission. Furthermore, based on the random waypoint (RWP) mobility model, the optimum update interval which provides maximum bidirectional user equipment (UE) throughput, has been derived.
Results show that the proposed schemes can achieve better average overall
throughput compared to the benchmark one-bit feedback and full-feedback
mechanisms. 
\end{abstract}
\vspace{-0.1cm}	
	
%\begin{IEEEkeywords}
%\vspace{-0.5cm}
%LiFi, Downlink, Uplink, Limited Feedback, Channel Update Interval
%\end{IEEEkeywords}\vspace{-0.4cm}

	\IEEEpeerreviewmaketitle

\section{Introduction}\vspace{-0.1cm}
\IEEEPARstart{T}{he} ever increasing number of mobile-connected devices, along with monthly global data traffic which is expected to be $35$ exabytes by $2020$ \cite{Cisco}, motivate both academia and industry to invest in alternative methods. These include mmWave, massive multiple-input multiple-output (MIMO), free space optical communication and Light-Fidelity (LiFi) for supporting future growing data traffic and next-generation high-speed wireless communication systems. Among these technol-\\ogies, LiFi is a novel bidirectional, high-speed and fully networked wireless communication technology. LiFi uses visible light as the propagation medium in downlink for the purposes of illumination and communication.  It may use infrared in uplink in order to not affect the illumination constraint of the room, and also not to cause interference with the visible light in the downlink \cite{Haas}. %It has been experimentally shown that $3.46$ Gb/s data rate over $5$ m free-space link can be achieved using visible light communication (VLC) with micro light emitting diode (LED) \cite{Islim} which is very promising for future indoor communications. 
LiFi offers considerable advantages in comparison to radio frequency (RF) systems. These include the very large, unregulated bandwidth available in the visible light spectrum, high energy efficiency, and rather straightforward deployment with off-the-shelf light emitting diode (LED) and photodiode (PD) devices at the transmitter and receiver ends respectively, enhanced security as the light does not penetrate through opaque objects \cite{WuVLC5G}. These notable benefits of LiFi have made it favourable for recent and future research.
	
It is known that utilizing channel adaptive signalling can bring on enhancement in almost any performance metric. Feedback can realize many kinds of channel adaptive methods that were considered impractical due to the problem of obtaining instantaneous channel state information (CSI) at the access point (AP). Studies have proven that permitting the receiver to transmit a small amount of information or feedback about the channel condition to the AP can provide near optimal performance \cite{LFchen2014performance, LFHeath1, LFHeath2,love2008overview}. Feedback conveys the channel condition, e.g., received power, signal-to-noise-plus-interference ratio (SINR), interference level, channel state, etc., and the AP can use the information for scheduling and resource allocation. The practical systems using this strategy, also known as limited-feedback (LF) systems, provide similar performance as the impractical systems with perfect CSI at the AP. 
	
It is often inefficient and impractical to continuously update the AP with the user equipment (UE) link condition. However, to support the mobility, it is also essential to consider the time-varying nature of channels for resource allocation problems to further enhance the spectral efficiency. With limited capacity, assignment of many resources to get CSI would evacuate the resources required to transmit actual data, resulting in reduced overall UE throughput \cite{jang2006throughput}. Therefore, it is common for practical wireless systems to update the CSI less frequently, e.g., only at the beginning of each frame. Many works have been done to reduce the amount of feedback in RF, however, very few studies are done to lessen the amount of feedback in optical wireless channels (OWCs). 
	
\vspace{-0.15cm}
\subsection{Literature Review and Motivation}\vspace{-0.15cm}
An overview of LF methods in wireless communications has been introduced in \cite{love2008overview}. The key role of LF in single-user and multi-user scenarios for narrowband, wideband communications with both single and multiple antennas has been discussed in \cite{love2008overview}. Two SINR-based limited-feedback scheduling algorithms for multi-user MIMO-OFDM in heterogeneous network is studied in \cite{pattanayak2016sinr} where UEs feed back channel quality information in the form of SINR. To reduce the amount of feedback, nearby UEs grouping and adjacent subcarrier clustering strategies have been considered.
In \cite{mokari2016limited}, three limited feedback resource allocation algorithms are evaluated for heterogeneous wireless networks. These resource allocation algorithms try to maximize the weighted sum of instantaneous data rates of all UEs over all cells. 
The authors in \cite{leinonen2008performance} proposed the ordered best-$K$ feedback method to reduce the amount of feedback. In this scheme, only the $K$ best resources are fed back to the AP.
	
An optimal strategy to transmit feedback based on outdated channel gain feedbacks and channel statistics for a single-user scenario has been proposed in \cite{vu2007capacity}. Other approaches are transmission of the quantized SINR of subcarriers which is the focus of \cite{sun2003asymptotic} and \cite{sun2003minimum}; and subcarrier clustering method which is developed in  \cite{agarwal2008multi} and \cite{svedman2004simplified}. In \cite{SaraScheduling},  the subcarrier clustering technique has been applied to the OWCs to reduce the amount of feedback by having each user send the AP the information of candidate clusters. %However, the feedback overhead depends on the number of clusters as it increases the feedback overhead increases.
A simple and more realizable solution, which is proposed in \cite{floren2003effect, gesbert2003selective, gesbert2004much}, is to inform the AP only if their SINR exceed some predetermined threshold. This is a very simple approach with only a one bit per subcarrier feedback. One-bit feedback method is very bandwidth efficient. However, using more feedback can provide slight downlink performance improvement but at the cost of uplink throughput degradation as discussed in \cite{floren2003effect}. The
benefits of employing only one bit feedback per subcarrier and the
minor data rate enhancements of downlink using more feedback bits are analyzed in \cite{sanayei2007opportunistic}. A one-bit feedback scheme for downlink OFDMA systems has been proposed in \cite{chen2006large}. It specifies whether the channel gain exceeds a predefined threshold or not. Then, UEs are assigned priority weights, and the optimal thresholds are chosen to maximize the weighted sum capacity. A problem linked to one-bit feedback technique is that there is a low probability that none of the UEs will report their SINR to the AP so that leaving the scheduler with no information about the channel condition. This issue can be solved at the expense of some extra feedback and overhead by the multiple-stage version of the threshold-based method proposed in \cite{hassel2007threshold}.  
	
The limited feedback approaches mentioned above are applicable to LiFi networks. However, due to fairly static behavior of LiFi channels, the feedback can be reduced further without any downlink throughput degradation.
	
\vspace{-0.2cm}
	
\subsection{Contributions and Outcomes}
In order to get the maximum bidirectional throughput, the amount of feedback should be optimized in terms of both quantity and update interval. In this paper, we proposed two methods to reduce the feedback information. The main contributions of this paper are outlined as follows.
	
$\bullet$ Proposing the modified carrier sense multiple access with collision avoidance (CSMA/CA) protocol suitable for uplink of LiFi networks.
	
$\bullet$ Proposing the limited-content feedback (LCF) scheme for LiFi networks which shows a close downlink performance to the full-feedback (FF) mechanism and even lower overhead compared to one-bit feedback technique.
	
$\bullet$ Proposing the limited-frequency feedback (LFF) scheme based on sum-throughput of uplink and downlink maximization. Deriving the optimum update interval for random waypoint (RWP) mobility model and investigating the effects of different parameters on it.
	
	%\subsection{Paper Organization}
	%The rest of this paper is organized as follows. The system model of  bidirectional LiFi networks is introduced in Section~\ref{sec2}. The downlink achievable throughput is calculated in Section~\ref{sec4}. In Section~\ref{sec5}, the modified CSMA/CA is proposed and the uplink throughput has been obtained. In Section~\ref{sec6}, the proposed LCF and LFF schemes are introduced and evaluated. Then, the optimum update interval is derived for the RWP mobility model. Finally, conclusions are drawn in Section~\ref{sec9}. 
	
	\vspace{-4pt}
	\section{System Model}
	\label{sec2}
	\subsection{Optical Attocell System Configuration}
\begin{figure}[t!]
		\centering
		\resizebox{0.65\linewidth}{!}{
			\includegraphics{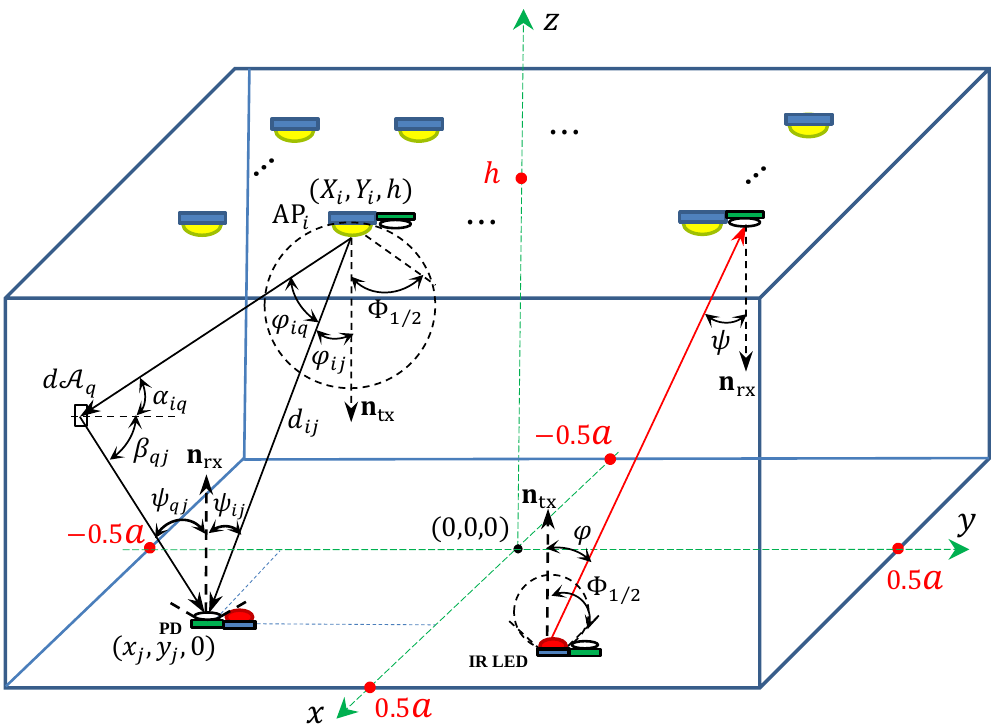}}\vspace{-0.2cm}  
		\caption{Geometry of light propagation in LiFi networks. Downlink (consist of LOS and NLOS components) and uplink (including LOS component) are shown with black and red lines, respectively.} 
		\label{fig1.2}
		\vspace{-1.3cm}  
	\end{figure}

	A bidirectional optical wireless communication has been considered in this study. In the downlink, visible light is utilized for the purpose of both illumination and communication, while in the uplink data are transmitted through infrared light in order not to affect the illumination constraint of the room. 
	The geometric configuration of the downlink/uplink in an indoor optical attocell network is shown in Fig.~\ref{fig1.2}. The system comprises multiple LED transmitters (i.e., APs) arranged on the vertexes of a square lattice over the ceiling of an indoor network and there is a PD receiver on UE. The LEDs are assumed to be point sources with Lambertian emission patterns. To avoid nonlinear distortion effects, the LEDs operate within the linear dynamic range of the current-to-power characteristic curve. In addition, the LEDs are assumed to be oriented vertically downwards, and the UE are orientated upward to the ceiling. Under this condition, the channel model for both downlink and uplink is the same. One AP is only selected to serve the UE based on the UE location. An optical attocell is then defined as the confined area on the UE plane in which an AP serves the UE. Frequency reuse (FR) plan is considered in both downlink and uplink to reduce the co-channel interference and also guarantee the cell edge users data rate. 
%	The room area is divided to $\Omega_{\rm{FR}}$ separate partitions, and $|A_n|$ is the area where $n$th portion of bandwidth is utilized. The allocated downlink bandwidth, $B_{{\rm{d}},n}$, and the allocated uplink bandwidth, $B_{{\rm{u}},n}$, in this area is given as:\vspace{0.cm}
%	\begin{equation}
%	\label{equation0}
%	\begin{aligned}
%	&B_{{\rm{d}},n}=B_{{\rm{d}}}\times \frac{|A_n|}{|A_{\rm{room}}|} \\
%	&B_{{\rm{u}},n}=B_{{\rm{u}}}\times \frac{|A_n|}{|A_{\rm{room}}|}, 
%	\end{aligned}
%	\end{equation}
%	for $n=1,2,3,4$; where $|A_{\rm{room}}|$ is the room area; $B_{{\rm{d}}}$ and $B_{{\rm{u}}}$ are the total downlink and uplink bandwidth, respectively. 
	Further details about FR plan can be found in \cite{HartmanPatent} and \cite{chen2015fractional}.
	
	Power- and frequency-based soft handover methods for visible light communication networks are proposed to reduce data rate fluctuations as the UE moves from one cell to another \cite{Dinc}.
	We consider power-based soft handover with the decision metric introduced in \cite{SoftHandoverChoi} as $|\gamma_{\imath} -\gamma_i|<\alpha$, where $\gamma_{\imath}$ and $\gamma_i$ are the SINR of the serving AP and adjacent APs, respectively; and $\alpha$ is the handover threshold. As a results the cell boundaries shape a circle with the radius of $r_{\rm{c}}$. According to the considered soft handover scheme, when the difference of SINR from two APs goes below the threshold, handover will occur. 
	
	The received optical signal at the PD consists of line of sight (LOS) and/or non-line of sight (NLOS) components. The LOS is a condition where the optical signal travels over the air directly from the transmitter to the UE, while the NLOS is a condition where the optical signal is received at the UE just by means of the reflectors. These two components are characterized as follows.
	\vspace{-10pt}
	
	\subsection{Light Propagation Model}
	The direct current (DC) gain of the LOS optical channel between the $i$th LED and the $j$th PD is given by:\vspace{-0.2cm}
	\begin{equation}
		\label{equation1}
		H_{{\rm{LOS}},{i,j}} \!=\! \begin{cases} \! \dfrac{(m+1)A}{2\pi d_{i j}^{2}}\cos ^{m}\!\phi_{i j}g_{\rm{f}}g(\psi_{i j})\cos\psi_{i j}, & 0\le \psi _{i j}\le \Psi_{c} \\
			\ 0 , & \psi _{i j}>\Psi_{\rm{c}} \end{cases},
	\end{equation}
	where $A$, $d_{ij}$, $\phi_{ij}$ and $\psi_{ij}$ are the physical area of the detector, the distance between the $i$th transmitter and the $j$th receiver surface, the angle of radiance with respect to the axis normal to the $i$th transmitter surface, and the angle of incidence with respect to the axis normal to the $j$th receiver surface, respectively. In \eqref{equation1}, $g_{\rm{f}}$ is the gain of the optical filter, and $\Psi_{\rm{c}}$ is the receiver field of view (FOV). In \eqref{equation1}, $g(\psi _{i}) =\varsigma^2 /\sin^2\Psi_{\rm{c}}$ for $0\le\psi_i\le\Psi_{\rm{c}}$, and $0$ for $\psi_{i}>\Psi_{\rm{c}}$, is the optical concentrator gain where $\varsigma$ is the refractive index; and also $m=-1/\log_2(\cos\Phi_{1/2})$ is the Lambertian order where $\Phi_{1/2}$ is the half-intensity angle \cite{ChenCheng}.
	The radiance angle $\phi_{ij}$ and the incidence angle $\psi_{ij}$ of the $i$th LED and $j$th UE are calculated using the rules from analytical geometry as $\cos\phi_{ij}={\bf{d}}_{ij}\cdot{\bf{n}}_{\rm{tx}}/{\Vert {\bf{d}}_{ij}\Vert }$ and $\cos\psi_{ij}=-{\bf{d}}_{ij}\cdot{\bf{n}}_{\rm{rx}}/{\Vert {\bf{d}}_{ij}\Vert }$, where ${\bf{n}}_{\rm{tx}}=[0, 0, -1]$ and ${\bf{n}}_{\rm{rx}}=[0, 0, 1]$ are the normal vectors at the transmitter and $j$th receiver planes, respectively and ${\bf{d}}_{ij}$ denotes the distance vector between $i$th LED and the $j$th UE and $\cdot$ and $\Vert \cdot\Vert$ denote the inner product and the Euclidean norm operators, respectively.
	
	In NLOS optical links, the transmitted signal arrives at the PD through multiple reflections. In practice, these reflections contain both specular and diffusive components. In order to keep a moderate level of analysis, first-order reflections only are considered in this study. A first-order reflection consists of two segments: i) from the LED to a small area $d\mathcal{A}_q$ on the wall; and ii) from the small area $d\mathcal{A}_q$ to the PD. The DC channel gain of the first-order reflections is given by:
	\vspace{-5pt}
	\begin{equation}
		\label{equation2}
		\begin{aligned}
			H_{{\rm{NLOS}},{i,j}} = \int_{\!\mathcal{A}_q}\!\!\frac{\!\rho_q(m\!+\!1)A}{2\pi^2d_{iq}^2d_{qj}^2} \cos^{m}\!\phi_{iq}\cos\psi_{qj}g_{f}g(\psi_{qj})\cos\alpha_{iq}\cos\beta_{qj} d\mathcal{A}_q ,
		\end{aligned}
	\end{equation}
where $\mathcal{A}_q$ denotes the total walls reflective area; $\rho_{q}$ is the reflection coefficient of the $q$th reflection element; $d_{iq}$ is the distance between the $i$th LED and the $q$th reflection element; $d_{qj}$ is the distance between the $q$th reflection element and the $j$th UE; $\phi_{iq}$ and $\psi_{iq}$ are the angle of radiance and the angle of incidence between the $i$th LED and the $q$th reflective element, respectively; and  $\phi_{qj}$ and $\psi_{qj}$ are the angle of radiance and the angle of incidence between the $q$th reflective element and the $j$th UE, respectively \cite{barry1993simulation}.
The channel gain between ${\rm{AP}}_i$ and ${\rm{UE}}_j$ is comprised of both LOS and NLOS components that is expressed as:\vspace{-0.2cm}
	\begin{equation}
		\label{equation}
		H_{i,j}=H_{{\rm{LOS}},i,j}+H_{{\rm{NLOS}},i,j}.
		\vspace{-0.2cm}
	\end{equation}
Note that due to symmetry of downlink and uplink channels, \eqref{equation1}-\eqref{equation} are valid for both downlink and uplink. %For $i=1,2,..,N_{\rm{AP}}$ and $j=1,2,..,N_{{\rm{UE}},i}$, where $N_{\rm{AP}}$ and $N_{{\rm{UE}},i}$ are the number of APs and UEs serving by ${\rm{AP}}_i$, respectively.
	\vspace{-10pt}
	
\subsection{Low Pass Characteristic of LED}
The frequency response of an off-the-shelf LED is not flat and is modeled as a first order low pass filter as, $H_{\rm{LED}}(w)=e^{-w/w_0}$, where $w_0$ is the fitted coefficient \cite{le2009100}. The higher the value of $w_0$, the wider the 3-dB bandwidth, $B_{\rm{3dB}}$. The 3-dB bandwidth of typical LEDs is low, however, the modulation bandwidth, $B$, can be multiple times greater than $B_{\rm{3dB}}$ thanks to utilization of OFDM. In this paper, we consider OFDMA for two purposes: \textit{i}) to alleviate the low pass effect of LED and \textit{ii}) to support multiple access. The frequency response of LED on $k$th subcarrier can be obtained as:\vspace{-0.3cm}
	\begin{equation}
		\label{equation3}
		H_{{\rm{LED}},k}=e^{-{2\pi kB_{{\rm{d}},n}}/{\mathcal{K}w_0}},
		\vspace{-0.2cm}
	\end{equation}
where $\mathcal{K}$ is the total number of subcarriers and $B_{{\rm{d}},n}$ is the downlink bandwidth of $n$th FR plan. 
	
\subsection{Receiver Mobility Model}
\begin{figure}[!t]
	\centering
	\subfloat[\label{sub1:14a}]{%
		\includegraphics[width=65mm,height=60mm]{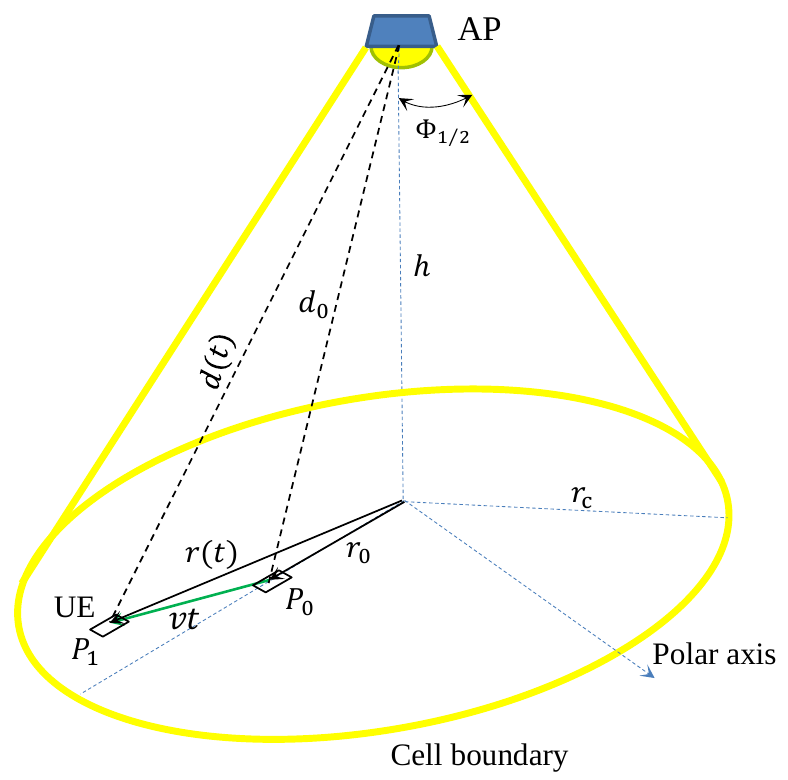}
	}
	\ 
	\subfloat[\label{sub2:14b}]{%
		\includegraphics[width=50mm,height=50mm]{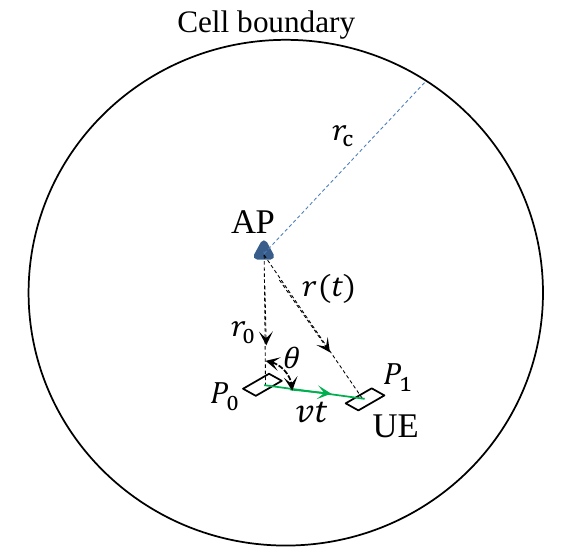}
	}\\
	\vspace{-0.2cm}
	\caption{RWP movement model.}
	\label{fig14}
	\vspace{-1cm}
\end{figure}

We considered the RWP model which is a commonly used mobility model for simulations of wireless communication networks \cite{Bettstetter}. The RWP mobility model is shown in Fig.~\ref{fig14}.
According to the RWP model, the UE's movement from one waypoint to another waypoint complies with a number of rules, including i) the random destinations or waypoints are chosen uniformly with probability $1/(\pi r_{\rm{c}}^2)$; ii) the movement path is a straight line; and iii) the speed is constant during the movement. The RWP mobility model can be mathematically expressed as an infinite sequence of triples: $\{({\bf{P}}_{\ell-1},{\bf{P}}_{\ell},v_{\ell})\}_{\ell\in\mathbb{N}}$ where $\ell$ denotes the $\ell$th movement period during which the UE moves between the current waypoint ${\bf{P}}_{\ell-1}\!=\!(x_{\ell-1},y_{\ell-1},0)$ and the next waypoint ${\bf{P}}_{\ell}=(x_\ell,y_\ell,0)$ with the constant velocity $V_{\ell}=v$. RWP model is more realistic scenario and has been used in many studies for modeling the mobility of UE \cite{Andrews}, \cite{EsaRWP}.

The UE distance at time instance $t$ from the AP is $d(t)=\left(r^2(t)+h^2\right)^{1/2}$, where  $r\!(t)=\!(r_0^2+v^2t^2-2r_0vt\cos\theta)^{1/2}$ with $\theta\!=\!\pi\!-\cos^{-1}\!\left(\!\frac{\boldsymbol{\vec{r_0}}\cdot \boldsymbol{\vec{v}}} {|\boldsymbol{\vec{r_0}}||\boldsymbol{\vec{v}}|}\! \right) $; $\boldsymbol{\vec{r_0}}$ is the initial UE distance vector from the cell center at $t\!=\!0$ with $|\boldsymbol{\vec{r_0}}|=r_0$; and $\boldsymbol{\vec{v}}$ is the vector of UE's velocity with $|\boldsymbol{\vec{v}}|=v$. Here, $r_0$ has the probability distribution function (PDF) of $f_{\mathcal{R}_0}(r_0)\!\!=\!\!{2r_0}/{r_{\rm{c}}^2}$ and $\theta$ is chosen randomly from a uniform distribution with PDF of $f_{\Theta}(\theta)\!\!=\!\!1/{\pi}$. For notation simplicity, the dependency of the equations to time is omitted unless it is confusing.

	\section{Downlink Throughput Calculation}
	\label{sec4}
The channel access protocol in the downlink is assumed to be orthogonal frequency division multiple access (OFDMA) based on DCO-OFDM so as to support downlink multiple access simultaneously. The modulated data symbols of different UEs, $X_k$, are arranged on $\mathcal{K}$ subcarriers of the OFDMA frame, $\boldsymbol{X}$. Then, the inverse fast Fourier transform (IFFT) is applied to the OFDMA frame to obtain the time domain signal $\tilde{x}$. For optical systems that perform intensity modulation, the modulated signal, $\tilde{x}$, must be both real and positive \cite{armstrong2008comparison}. This requires two constraints on the entities of OFDMA frame: i) $\boldsymbol{X}(0)=\boldsymbol{X}(\mathcal{K}/2)=0$, and ii) the Hermitian symmetry constraint, i.e., $\boldsymbol{X}(k)=\boldsymbol{X}^*(\mathcal{K}-k)$, for $k\neq0$, where $(\cdot)^*$ denotes the complex conjugate operator. Therefore, the OFDMA frame is  $\boldsymbol{X}=\zeta[0, X_1, ..., X_{\mathcal{K}/2-1}, 0, X^*_{\mathcal{K}/2-1},..., X^*_1]$, the normalizing factor, $\zeta=\sqrt{\mathcal{K}/(\mathcal{K}-2)}$, is multiplied since the $0$th and $(\mathcal{K}/2)$th samples require no energy. Note that the number of modulated subcarriers bearing information is $\mathcal{K}/2-1$. Afterwards, a moderate bias relative to the standard deviation of the AC signal $\tilde{x}$ is used as $x_{\rm{DC}}=\eta\sqrt{\mathbb{E}[\tilde{x}^2]}$ \cite{dissanayake2013comparison}. The signal $x=x_{\rm{DC}}+\tilde{x}$ is then used as the input of an optical modulator.

Let $\boldsymbol{H}_j=[H_{i,j}]$, for $i=1,2, ..., N_{\rm{AP}}$, be the downlink visible light channel gain vector from all APs to the ${\rm{UE}}_j$. The ${\rm{UE}}_j$ is connected to ${\rm{AP}}_{\imath}$ based on the maximum channel gain criterion so that $\imath=\arg_i\max({\bf{H}}_j)$. Afterwards, the embedded scheduler algorithm in ${\rm{AP}}_{\imath}$ allocates a number of subcarriers to the ${\rm{UE}}_j$ based on its requested data rate and its link quality. 
In this study, a fair scheduling method for OFDMA-based wireless systems is considered \cite{jalali2000data,kim2002proportionally}. The scheduler assigns the $k$th resource to $j$th UE according to the following metric:\vspace{-0.1cm}
\begin{equation}
\label{equation4}
j=\arg \max_i \frac{R_{{\rm{req}},j}}{\overline{R_i}},
\end{equation}
where $\overline{R_i}$ is the average data rate of $i$th UE before allocating the $k$th resource, and $R_{{\rm{req}},j}$ is the request data rate of ${\rm{UE}}_j$.
	
Throughout this study, we consider LiFi systems transmitting data based on DC-biased optical OFDM, for which the upper bound on the achievable data rate can be expressed in a Shannon capacity expression form as a function of \textit{electrical} SINR as shown in \cite{dimitrov2013information}. Assume the effect of clipping noise is negligible, the downlink rate of ${\rm{UE}}_j$ after scheduling can be obtained as:\vspace{-0.0cm}
\begin{equation}
\label{equation5}
R_{{\rm{d}},j}=\dfrac{B_{{\rm{d}},n}}{\mathcal{K}}\sum_{k=1}^{\mathcal{K}/2-1}\log_2\left( 1+s_{j,k}\gamma_{{\rm{d}},j,k}\right), 
\vspace{-0.05cm}
\end{equation} 
where $s_{j,k}\!=\!1$ if $k$th subcarrier is allocated to the ${\rm{UE}}_j$ otherwise $s_{j,k}\!=\!0$; $\gamma_{{\rm{d}},j,k}$ is the SINR of ${\rm{UE}}_j$ on $k$th subcarrier serving by ${\rm{AP}}_{\imath}$. In communication systems, SINR is defined as the ratio of the desired electrical signal power to the total noise and interference power and is an important metric to evaluate the connection quality and the transmission data rate. Denoting $P_{{\rm{elec}},\imath,j,k}$ as the received electrical power of $j$th UE on $k$th subcarrier, then, $\gamma_{{\rm{d}},j,k}= {P_{{\rm{elec}},\imath,j,k}}/ {(\sigma_{j,k}^2+\!P_{{\rm{int}},j})}$, where $\sigma_{j,k}^2\!=\!N_0B_{{\rm{d}},n}/\mathcal{K}$, is the noise on $k$th subcarrier of ${\rm{UE}}_j$, and $N_0$ is the noise power spectral density; $P_{{\rm{int}},j}$ is the interference from other APs on $j$th UE. It is assumed that the APs emit the same average optical power and the total transmitted electrical power is equally allocated among $\mathcal{K}-2$ subcarriers so that the received electrical power on $k$th subcarrier of $j$th UE is equal to  $P_{{\rm{elec}},\imath,j,k}=R_{\rm{PD}}^2P_{{\rm{d,opt}}}^2H_{\imath,j,k}^2H_{{\rm{LED}},k}^2/(\eta^2(\mathcal{K}-2))$, where $P_{{\rm{d,opt}}}$ is the transmitted optical power; $R_{\rm{PD}}$ and $\eta$ are the PD responsivity and conversion factor, respectively; $H_{\imath,j,k}$ is the frequency response of channel gain on $k$th subcarrier. It includes both LOS and the first order reflections. Accordingly, the received SINR of $j$th UE on $k$th subcarrier can be expressed as:\vspace{-0.0cm}
\begin{equation}
\label{equation6}
\begin{aligned}
&\gamma_{{\rm{d}},j,k}\!=\!\dfrac{R_{\rm{PD}}^2P_{\rm{d,opt}}^2 H_{\imath,j,k}^2H_{{\rm{LED}},k}^2} {(\mathcal{K}\!-\!2)\eta^2\sigma_{j,k}^2\!+\!\!\!\!\!\!\sum\limits_{i\in \mathcal{S}_{\rm{AP},\imath}}\!\!\!\!\!\!R_{\rm{PD}} ^2P_{\rm{d,opt}}^2 H_{i,j,k}^2H_{{\rm{LED}},k}^2}.  
\end{aligned}
\end{equation}\vspace{-0.05cm}
where $\mathcal{S}_{\rm{AP},\imath}$ is the set of all other APs using the same frequencies as the ${\rm{AP}}_{\imath}$.

\section{Uplink Throughput Calculation}
\label{sec5}
\subsection{Uplink Access Protocol}
In this study, CSMA/CA is considered as the uplink access protocol. CSMA/CA is a multiple access protocol with a binary slotted exponential backoff strategy being used in wireless local area networks (WLANs) \cite{IEEEMACStand}. This is known as the collision avoidance mechanism of the protocol. In CSMA/CA, a UE will access the channel when it has data to transmit. Thus, this access protocol uses the available resources efficiently. Once the UE is allowed to access the channel, it can use the whole bandwidth. 
However, this access protocol cannot directly be used in LiFi networks, because it results in severe ``hidden node" problem. Here, we applied two simple modifications to CSMA/CA to minimize the number of collisions in LiFi networks. Firstly, the request-to-send/clear-to-send (RTS/CTS) packet transmission scheme, which is optional in WLANs should be mandatory in LiFi networks. This is the only way that UEs can notice that the channel is busy in LiFi networks. The reason behind this is that different wavelengths are employed in downlink and uplink of LiFi networks, visible light and infrared, respectively. Thus, the PD at the UE is tuned for visible light and cannot sense the channel when another UE transmits via infrared. Secondly, the AP transmits a channel busy (CB) tone to inform the other UEs that the channel is busy. In the following, the modified CSMA/CA is described in detail.  	\vspace{-.2cm}
%	\begin{figure*}[!Ht]
%		\centering
%		\resizebox{0.8\linewidth}{!}{
%			\includegraphics{Fig4.pdf}}
%		\caption{Conventional and modified four-way handshaking RTS/CTS mechanism} 
%		\label{figu3}  
%		\vspace{-1cm}
%	\end{figure*}
\subsection{Brief Description of the Access Protocol}
In CSMA/CA, UEs listen to the channel prior to transmission for an interval called distributed inter-frame space (DIFS). Then, if the channel is found to be idle, the UEs generate a random backoff, $\mathcal{B}_j$, for $j=1,2,\ldots,N$, where $N$ is the number of competing UEs.  
The value of $\mathcal{B}_j$ is uniformly chosen in the range $[0, w-1]$, where $w$ is the contention window size. Let $\boldsymbol{\mathcal{B}}=[\mathcal{B}_j]_{1\times N}$, be the backoff vector of the UEs. After sensing the channel for time interval DIFS, ${\rm{UE}}_j$ should wait for $\mathcal{B}_j\times t_{\rm{slot}}$ seconds, where $t_{\rm{slot}}$ is the duration of each time slot. Obviously, the UE with lowest backoff is prior to transmit, i.e., $u_1$th UE, where $u_1=\arg_j\min(\boldsymbol{\mathcal{B}})$. Then, $u_1$th UE sends the RTS frame to the AP before $N-1$ other UEs. If the RTS frame received at the AP successfully, it replies after a short inter-frame space (SIFS) with the CTS frame. The $u_1$th UE only proceeds to transmit the data frame, after the time interval of SIFS, if it receives the CTS frame. Eventually, an ACK is transmitted after the period of SIFS by the AP to notify the successful packet reception. The AP transmits the CB tone simultaneously with the reception of RTS packet. The UEs that can hear the CB tone will freeze their backoff counter. The backoff counter will reactivated when the channel is sensed idle again after the period of DIFS. If the AP does not transmit the CB tone, the $u_2$th UE who cannot hear the $u_1$th UE, will start to send RTS frame after waiting for $\mathcal{B}_{u_2}\times t_{\rm{slot}}$ seconds. Here, $u_2$th UE is called the hidden UE and a collision occurs if $(\mathcal{B}_{u_2}-\mathcal{B}_{u_1})\times t_{\rm{slot}}<t_{\rm{RTS}}$, where $t_{\rm{RTS}}$ is the RTS frame transmission time which is directly proportional to the length of RTS frame, $L_{\rm{RTS}}$, and inversely proportional to the uplink rate.  
	
%	RTS and CTS frames carry a network allocation vector (NAV) field which is a virtual carrier sensing mechanism. NAV indicates the number of microseconds that the channel is reserved by a UE. The NAV in RTS frame includes the CTS frame, the data frame, and the subsequent ACK frame from the AP. The CTS frame contains a new NAV updated to the time already elapsed. After the CTS frame is sent, all UEs that can receive the CTS will update their NAV timer and defer transmission until their NAV timer reach zero. This keeps the channel free for the $u_1$th UE to complete the data transmission process and alleviates the problem of hidden UEs.
	
\subsection{Uplink Throughput}
In the modified CSMA/CA for LiFi networks, collision only occurs if the backoff time of at least two UEs reach to zero simultaneously. Thus, they transmit at the same time and the packets collide. The analysis of normalized throughput and collision probability is the same as the analysis provided in \cite{bianchi2000performance}. In the following, we only provide a summary of the equations and further detail is provided in \cite{bianchi2000performance}. The normalized uplink throughput is given as:\vspace{0.cm}
\begin{equation}
\label{equation13}
\begin{aligned}
&\widetilde{\mathcal{T}}_{\rm{u}}=\dfrac{P_{\rm{t}}P_{\rm{s}}\mathbb{E}[t_{\rm{D}}]}{(1-P_{\rm{t}})t_{\rm{slot}}+P_{\rm{t}}P_{\rm{s}}\mathbb{E}[t_{\rm{s}}]+P_{\rm{t}}(1-P_{\rm{s}})t_{\rm{c}}},\\
\end{aligned}
\end{equation}
where $P_{\rm{t}}=1-(1-\tau)^{N}$ is the probability of at least one transmission in the considered backoff slot time, $P_{\rm{s}}={N\tau(1-\tau)^{N-1}}/{P_{\rm{t}}}$ is the probability of successful transmission, and $\tau=\frac{2}{w+1}$ is the probability that a UE transmits on a randomly chosen slot time.
In \eqref{equation13}, $\mathbb{E}[t_{\rm{D}}], \mathbb{E}[t_{\rm{s}}]$ and $t_{\rm{c}}$ are the average transmission time of data packet, average successful transmission time and collision time, respectively. Assuming that all data packets have the same length, then:
\vspace{-1pt}
\begin{equation}
\label{equation14}
\begin{aligned}
\mathbb{E}[t_{\rm{s}}]&=t_{\rm{s}}=t_{\rm{RTS}}+{\rm{SIFS}}+t_{\rm{dely}}+t_{\rm{CTS}}+{\rm{SIFS}}+t_{\rm{dely}}+t_{\rm{HDR}} +t_{\rm{D}}+{\rm{SIFS}}\\\vspace{-0.1cm}&+t_{\rm{dely}}+t_{\rm{ACK}} +{\rm{DIFS}}+t_{\rm{dely}}, \ \ \  \mathbb{E}[t_{\rm{D}}]=t_{\rm{D}}, \ \ \  t_{\rm{c}}=t_{\rm{RTS}}+{\rm{DIFS}}+t_{\rm{dely}}
\end{aligned}
\end{equation}\vspace{-0.1cm}
where $t_{\rm{dely}}$ is the propagation delay. Note that the packet header includes both physical and MAC header. 
Finally, the uplink throughput of $j$th UE can be obtained as follows:
\vspace{-0.1cm}
\begin{equation}
\label{equation15}
R_{{\rm{u}},j}=\frac{\widetilde{\mathcal{T}}_{\rm{u}}B_{{\rm{u}},n}}{N}\log_2\left(1+\gamma_{{\rm{u}},j} \right).
\end{equation}
where $B_{{\rm{u}},n}$ is the uplink bandwidth of $n$th FR plan and $\gamma_{{\rm{u}},j}$ is the SINR at the AP when communicating with ${\rm{UE}}_j$ and it is given as: \vspace{-0.1cm}
\begin{equation}
\label{equation11}
\gamma_{{\rm{u}},j}=\dfrac{\left(R_{\rm{PD}}P_{\rm{u,opt}}H_{\imath,j}\right)^2} {\eta^2N_0B_{{\rm{u}},n}+\sum_{j\in \Pi}\left(R_{\rm{PD}}P_{\rm{u,opt}}H_{i,j}\right)^2},
\end{equation}
where $\Pi$ is the set of other UEs using the same bandwidth as ${\rm{UE}}_j$ and communicating with $i$th AP, $(i\neq\imath)$, simultaneously with ${\rm{UE}}_j$; and $P_{\rm{u,opt}}$ is the transmitted uplink power which is assumed to be the same for all UEs.\vspace{-0.cm}
	
\section{Feedback Mechanism}
\label{sec6}
\begin{figure}[t!]
\centering
\resizebox{0.68\linewidth}{!}{
\includegraphics{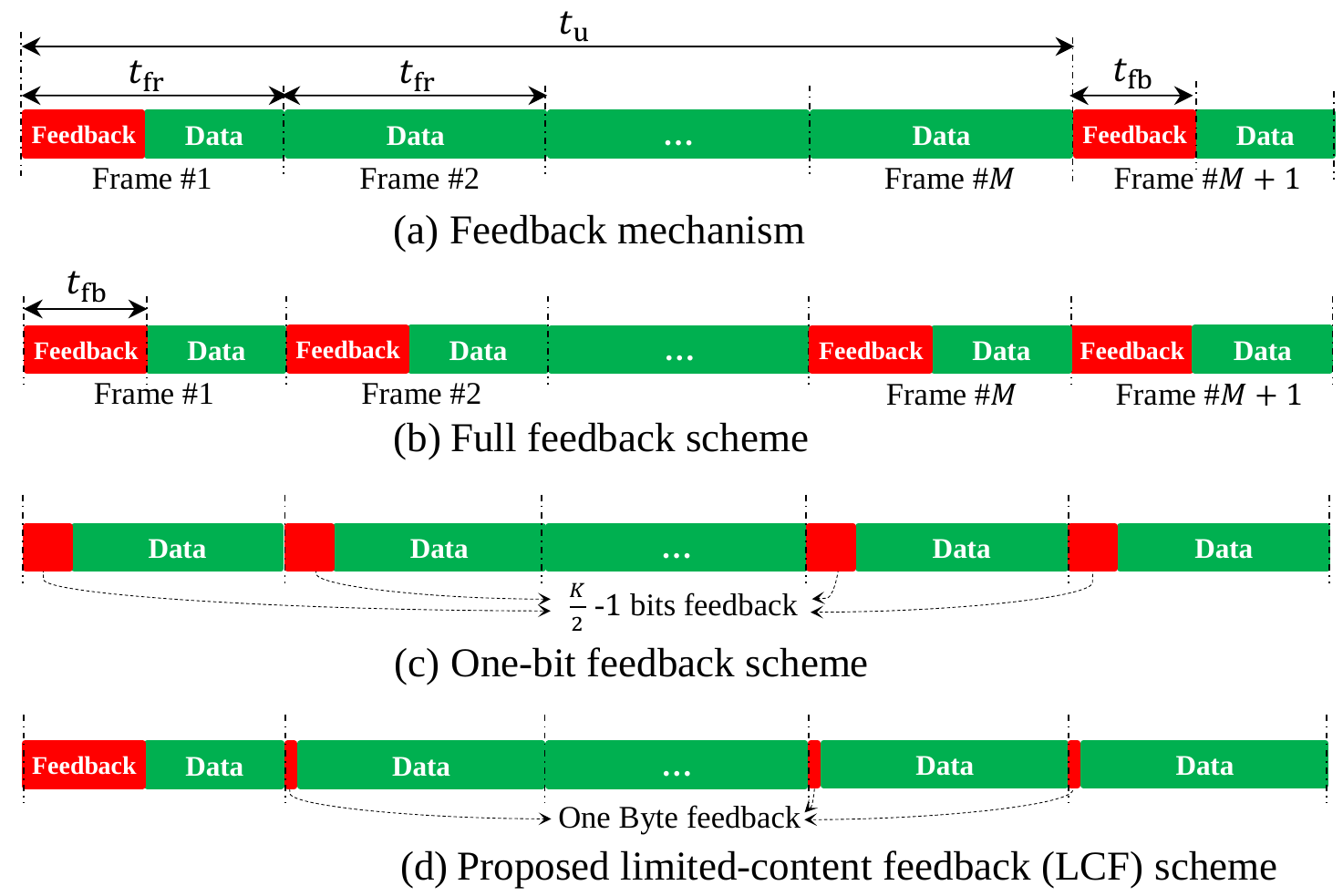}}
%\vspace{-0.6cm}
\caption{Feedback schemes.}  
\label{figu7} 
\vspace{-1.4cm}
\end{figure}

Over the last few years, studies have repeatedly illustrated that permitting the receiver to send some information bits about the channel conditions to the transmitter can allow effective resource allocation and downlink throughput enhancement. This feedback information is usually the SINR of a subcarrier at the receiver \cite{love2008overview, mokari2016limited}. However, sending this information is in cost of uplink throughput degradation. Therefore, there is a trade-off between downlink and uplink throughput when the amount of feedback varies.
Let's define the feedback ratio, $\epsilon$, as the ratio of total feedback time and total transmission time as:\vspace{-0.0cm}
\begin{equation}
\label{equation13_1}
\epsilon=\frac{\sum t_{\rm{fb}}}{t_{\rm{tot}}},
\end{equation}
where $t_{\rm{fb}}$ is the feedback duration. 
where $t_{\rm{fb}}$ is the feedback duration. 
Fig.~\ref{figu7}-(a) denotes a general feedback mechanism, in which feedback information is transmitted periodically after an interval of $t_{\rm{u}}$. Denoting that the denominator of \eqref{equation13_1} is the total transmission time which is equal to $t_{\rm{tot}}=(N_{\rm{D}}+N_{\rm{f}})t_{\rm{fr}}$, where $N_{\rm{D}}$ and $N_{\rm{f}}$ are the number of data and feedback frames in the total transmission time. The total feedback time is $\Sigma t_{\rm{fb}}=N_{\rm{f}}t_{\rm{fb}}$. Replacing these equations in \eqref{equation13_1}, the feedback ratio can be obtained as:\vspace{-0.0cm}
\begin{equation}
\label{equation28}
\epsilon=\frac{N_{\rm{f}}t_{\rm{fb}}}{(N_{\rm{D}}+N_{\rm{f}})t_{\rm{fr}}}=\frac{t_{\rm{fb}}}{\left(1+\frac{N_{\rm{D}}}{N_{\rm{f}}}\right) t_{\rm{fr}}}.
\end{equation}
Since $t_{\rm{tot}}=(N_{\rm{D}}+N_{\rm{f}})t_{\rm{fr}}=N_{\rm{f}}t_{\rm{u}}$, then $1+\dfrac{N_{\rm{D}}}{N_{\rm{f}}}=\dfrac{t_{\rm{u}}}{t_{\rm{fr}}}$, and substituting it in \eqref{equation28}, it can be simplified as:\vspace{-0.3cm}
\begin{equation}
\label{equation16_1}
\epsilon=\dfrac{t_{\rm{fb}}}{t_{\rm{u}}}.
\vspace{-0.1cm}
\end{equation}
Then, the uplink throughput of ${\rm{UE}}_j$ in consideration of feedback is given by: \vspace{0.cm}
\begin{equation}
\label{equation16}
R_{{\rm{u}},j}=\left( 1-\dfrac{t_{\rm{fb}}}{t_{\rm{u}}}\right)  \frac{\widetilde{\mathcal{T}}_{\rm{u}}B_{{\rm{u}},n}}{N}\log_2\left(1+\gamma_{\rm{u},j} \right).
\end{equation}
	
Due to the use of DCO-OFDM modulation, the AP requires the SINR information of $\mathcal{K}/2-1$ subcarriers. The extreme and least cases for sending the SINR information are full feedback (FF) and one-bit fixed-rate feedback, respectively. These schemes are shown in Fig.~\ref{figu7}-(b) and Fig.~\ref{figu7}-(c). In the FF scheme, UEs send the SINR of all subcarriers at the beginning of each data frame. Obviously, this impractical method produces huge amount of feedback. According to one-bit feedback technique, the AP sets a threshold for all UEs. Each UE compares  the value of its SINR to this threshold. When the SINR exceeds the threshold, a `$1$' will be transmitted to the AP; otherwise a `$0$' will be sent. The AP receives feedback from all UEs and then randomly selects a UE whose feedback bit was `$1$'. If all the feedback bits received by the AP are zero, then no signal is transmitted in the next time interval. However, in this case the AP can also randomly chooses a UE for data transmission, although for large number of UEs this method has vanishing benefit over no data transmission when all the received feedback bits are `$0$' \cite{sanayei2005exploiting}.
	
As can be induced from \eqref{equation16_1}, the feedback ratio can generally be reduced by means of either decreasing the content of feedback or increasing the update interval. In the following, we propose the limited-content feedback (LCF) and limited-frequency feedback (LFF) techniques. The former is based on reducing the feedback information in each frame and the latter is based on increasing the update interval.
	
\subsection{Proposed Limited-content feedback (LCF) Scheme}
\begin{figure}[t]
\centering
\resizebox{0.55\linewidth}{!}{
\includegraphics{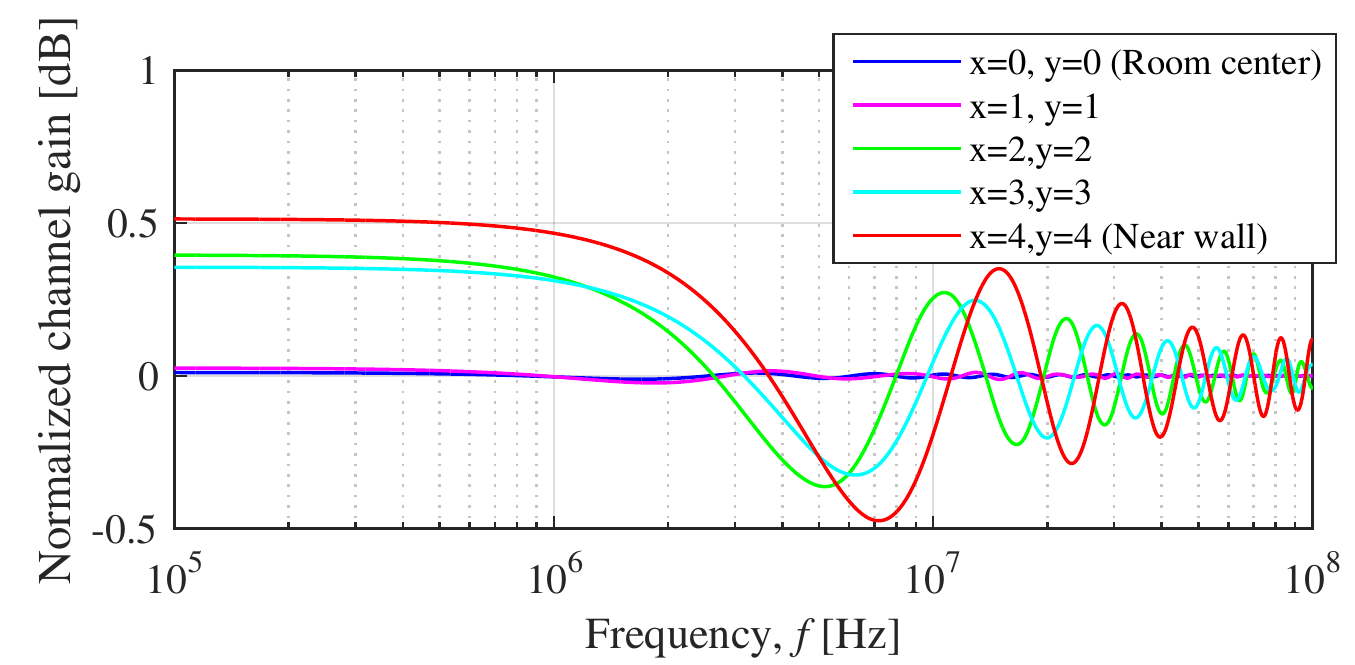}}\vspace{-0.2cm}
\caption{Normalized channel gain, $\frac{|H_{\imath,j}(f)|^2}{|H_{{\rm{LOS}},\imath,j}(f)|^2}$, for different room positions.} 
\label{figu2}
\vspace{-.3cm}  
\end{figure}

Unlike RF wireless and optical diffused channels, the frequency selectivity of the channel in LiFi attocell networks is mostly characterized by the limitations of the receiver/transmitter devices (i.e., PD and LED), rather than the multipath nature of the channel \cite{ChenCheng}. In order to assess the frequency response of the free-space optical channels, computer simulations are conducted. The simulations are performed for a network size of $10\times10\times2.15$ m$^3$. The network area is divided equally into nine quadrants with one AP located at the center of each. Assume the center of $xy$-plane is located in the center of the room as shown in Fig.~\ref{fig1.2}. The other parameters are listed in Table~\ref{TableI}. The normalized frequency response of the channel gain, $\frac{|H_{i,j}(f)|^2}{|H_{{\rm{LOS}},i,j}(f)|^2}$, for a UE placed at different positions of the room is depicted in Fig.~\ref{figu2}. As can be seen, the normalized frequency response fluctuates around the LOS component and the variation of the fluctuation is less than 1 dB. Moreover, the channel gain variation is less significant for UEs that are further away from the walls of the room, due to the lower significance of the first order reflection \cite{chen2015fractional}.
	\begin{table}[t]
		\centering
		\caption{Simulation Parameters}
		\label{TableI}
		\vspace{-8pt}
		{\raggedright
			\vspace{4pt} \noindent
			\begin{tabular}{p{130pt}|p{30pt}|p{80pt}}
				\hline
				\parbox{110pt}{\centering{\small Parameter}} & \parbox{30pt}{\centering{\small Symbol}} & \parbox{80pt}{\centering{\small Value}} \\
				\hline
				\hline
				\parbox{110pt}{\raggedright{\small Network space}} & \parbox{25pt}{\centering{\small --}} & \parbox{80pt}{\centering{\small $10\times10\times2.15$ $\text{m}^3$}} \\
				\hline
				\parbox{110pt}{\raggedright{\small Number of APs}} & \parbox{25pt}{\centering{\small $N_{\rm{AP}}$}} & \parbox{80pt}{\centering{\small $9$}} \\
				\hline
				\parbox{110pt}{\raggedright{\small Cell radius}} & \parbox{25pt}{\centering{\small $r_{\rm{c}}$}} & \parbox{80pt}{\centering{\small $2.35$ m}} \\
				\hline
				\parbox{110pt}{\raggedright{\small LED half-intensity angle}} & \parbox{25pt}{\centering{\small $\Phi_{1/2}$}} & \parbox{70pt}{\centering{\small $60^\circ$}} \\
				\hline
				\parbox{110pt}{\raggedright{\small Receiver FOV}} & \parbox{25pt}{\centering{\small $\Psi_{\rm{c}}$}} & \parbox{70pt}{\centering{\small $90^\circ$}} \\
				\hline
				\parbox{115pt}{\raggedright{\small Physical area of a PD}} & \parbox{25pt}{\centering{\small $A$}} & \parbox{65pt}{\centering{\small $1$ cm$^2$}} \\
				\hline
				\parbox{110pt}{\raggedright{\small Gain of optical filter}} & \parbox{25pt}{\centering{\small $g_{\rm{f}}$}} & \parbox{65pt}{\centering{\small $1$}} \\
				\hline
				\parbox{110pt}{\raggedright{\small Refractive index}} & \parbox{25pt}{\centering{\small $\varsigma$}} & \parbox{65pt}{\centering{\small $1$}} \\
				\hline
				\parbox{110pt}{\raggedright{\small PD responsivity}} & \parbox{25pt}{\centering{\small $R_{\rm{PD}}$}} & \parbox{65pt}{\centering{\small $1$ A/W }} \\
				\hline
				\parbox{110pt}{\raggedright{\small Reflection coefficient}} & \parbox{25pt}{\centering{\small $\rho_q$}} & \parbox{65pt}{\centering{\small $0.85$  }} \\
				\hline
				\parbox{110pt}{\raggedright{\small Number of subcarriers}} & \parbox{25pt}{\centering{\small $\mathcal{K}$}} & \parbox{65pt}{\centering{\small $2048$}} \\
				\hline
				\parbox{110pt}{\raggedright{\small Transmitted optical power}} & \parbox{25pt}{\centering{\small $P_{\rm{d,opt}}$}} & \parbox{65pt}{\centering{\small $8$ W}} \\
				\hline
				\parbox{110pt}{\raggedright{\small Downlink FR bandwidth}} & \parbox{25pt}{\centering{\small $B_{{\rm{d}},n}$}} & \parbox{65pt}{\centering{\small $10$ MHz}} \\
				\hline
				\parbox{110pt}{\raggedright{\small Fitted coefficient}} & \parbox{25pt}{\centering{\small $w_0$}} & \parbox{65pt}{\centering{\small $45.3$ Mrad/s}} \\
				\hline
				\parbox{110pt}{\raggedright{\small Conversion factor}} & \parbox{25pt}{\centering{\small $\eta$}} & \parbox{65pt}{\centering{\small $3$ }} \\
				\hline
				\parbox{130pt}{\raggedright{\small Noise power spectral density}} & \parbox{25pt}{\centering{\small $N_0$}} & \parbox{65pt}{\centering{\small $10^{-21}$ A$^2$/Hz }} \\
				\hline
			\end{tabular}
			\vspace{2pt}
		}
		\vspace{-1cm}
	\end{table}
Accordingly, the frequency selectivity of LiFi channels is mainly confined by LED and PD components, and the frequency selectivity of these devices are fairly static. The average received power at the UE is much more dynamic and is significantly dependent on the position of the UE. Therefore, by only updating the average power, a reasonable estimate of the SINR of all the subcarriers can be obtained.
This idea forms the foundation of our LCF scheme. 

Fig.~\ref{figu7}-(d) represents the principal working mechanism of our proposed LCF scheme. According to the LCF scheme, when a UE connects to an AP, it sends the SINR information of all subcarriers only once at the beginning of the first frame. For the following frames and as long as  the UE is connected to the same AP, it only updates the scheduler on its received average power (i.e., the DC channel component). Once the UE connects to a new AP, it will transmit the SINR information of all subcarriers again.
The proposed LCF scheme then simply scales the individual SINR values received in the next frames such that the total average power matches the updated average power \cite{DehghaniSoltani2015}. Thus, the estimated SINR on $k$th subcarrier of $j$th UE at time instance $t$ is given as:\vspace{-0.1cm}
\begin{equation}
\label{equation26}
\hat{\gamma }_{{\rm{d}},j,k}(t)\approx\gamma_{{\rm{d}},j,k}(0)\times \frac{\gamma_{{\rm{d}},j,0}(t)}{\gamma_{{\rm{d}},j,0}(0)}, 
\end{equation}
where $\gamma_{{\rm{d}},j,k}(0)$ is the downlink SINR of $j$th UE on $k$th subcarrier at $t=0$. The scheduler uses this estimated SINR information for subcarrier allocation according to \eqref{equation4}.
%In order to have an accurate estimation of SINR on each subcarrier, it is assumed that the UEs transmit the SINR of each subcarriers every $t_{\rm{u}}$ seconds as shown in Fig.~\ref{figu6}-(c). 
%It is worth mentioning that due to producing high amount of overhead, FF scheme is impractical but in order to show the close performance of proposed LF and FF techniques, FF scheme is considered in the simulation results.
	
\begin{figure}[t!]
\centering
\resizebox{0.5\linewidth}{!}{
\includegraphics{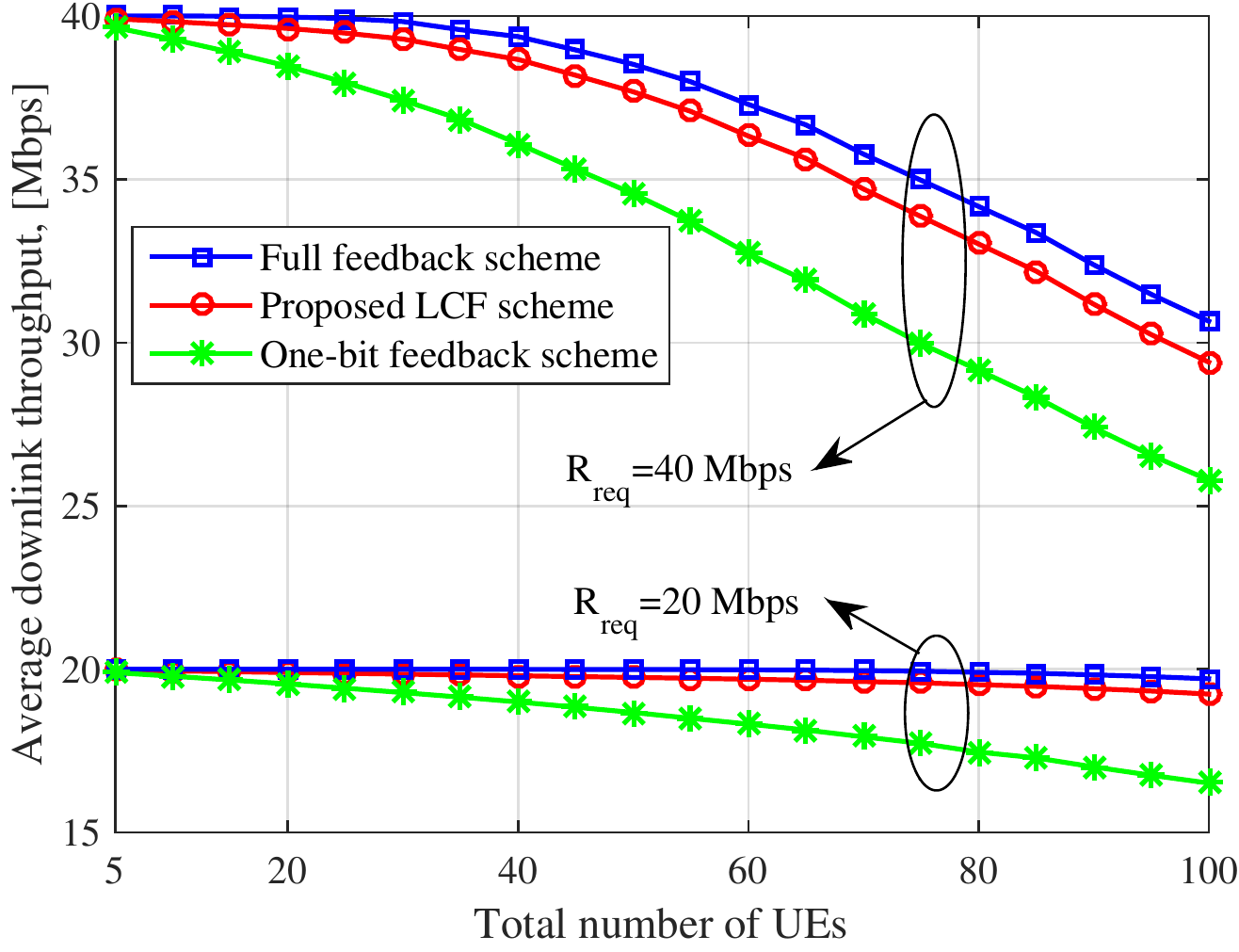}}\vspace{-0.2cm}
\caption{Average downlink throughput for different feedback schemes (average request data rate: 20 Mbps and 40 Mbps).} 
\label{fig4}
\vspace{-1cm}  
\end{figure}

The most salient difference between the LCF technique and one-bit feedback method is that the AP does not have any knowledge about the SINR value of each subcarrier and it just knows that the SINR is above or lower than a predetermined threshold for one-bit feedback technique. However, thanks to the use of LCF approach, the AP can have an estimation of the SINR value for each subcarrier. 
In order to compare the downlink performance of FF, one-bit feedback and LCF, Monte-Carlo simulations are executed. The simulation tests are carried out $10^3$ times per various number of UEs, and with the parameters given in Table~\ref{TableI}. In each run, the UEs' locations are chosen uniformly random in the room. Once they settle in the new locations, they update the AP about their subcarrier SINR as explained. Then, the AP, reschedule the resources based on \eqref{equation4}. The request data rate of UEs are assumed to be the same.
Fig.~\ref{fig4} illustrates the average downlink throughput versus different number of UEs for LCF, FF and on-bit feedback schemes. As can be seen from the results, the performance of the LCF is better than the one-bit feedback scheme and nearly similar to FF scheme. As the number of UEs increase the gap between the considered feedback schemes also increases. However, the LCF follows the FF fairly good especially for low data request rate. Moreover, compared to the one-bit feedback technique, the LCF scheme occupies less portion of the uplink bandwidth.\vspace{-0.2cm}  
	
	
\subsection{Proposed Limited-frequency feedback (LFF) Scheme}
Due to fairly static feature of LiFi channels, the UE can update the AP about its channel condition less frequently, especially when the UE is immobile or it moves slowly \cite{MDSPIMRC2017}.  
%The channel update interval, $t_{\rm{u}}$, is defined as the time period between two consecutive feedback transmission from the UE to its serving AP. 
%As shown in Fig.~\ref{figu7}, after transmission of $M$ frames, $(M+1)$th frame is assigned for feedback transmission to the AP. Hence the update interval is $t_{\rm{u}}=Mt_{\rm{fr}}$, where $t_{\rm{fr}}$ is the duration of one frame including the RTS/CTS access protocol header\footnote{This header has been once considered in the normalized factor $\widetilde{\mathcal{T}}_{\rm{u}}$.}; and $M\in \mathbb{Z}^{+}$ is a positive integer number. 
Based on the information of UE's velocity, we aim to find the appropriate channel update interval, $t_{\rm{u}}$, so that the expected weighted average sum throughput of uplink and downlink per user is maximized. Weighted sum throughput maximization is commonly used to optimize the overall throughput for bidirectional communications \cite{cirik2015weighted}, \cite{aquilina2017weighted}. The optimization problem (OP) is formulated as:\vspace{0.cm}
\begin{equation}
\label{equation17}
\begin{aligned}
\max_{t_{\rm{u}}} \left(\mathbb{E}_{\boldsymbol{[r_0]},\boldsymbol{[\theta]}}\left[ \frac{1}{N} \sum_{j=1}^{N}\left( w_{\rm{d}}\overline{R}_{{\rm{d}},j}(t_{\rm{u}})+w_{\rm{u}}\overline{R}_{{\rm{u}},j}(t_{\rm{u}})\right) \right]  \right),
\end{aligned}
\end{equation}
where $\overline{R}_{{\rm{d}},j}$ and $\overline{R}_{{\rm{u}},j}$ are the average downlink and uplink throughput of $j$th UE, respectively; Note that  $\boldsymbol{[r_0]}=[{r_0}_1,\cdots,{r_0}_N]$ and $\boldsymbol{[\theta]}=[\theta_1,\cdots,\theta_N]$ are random variable vectors with i.i.d entities; $\mathbb{E}_{\boldsymbol{[r_0]},\boldsymbol{[\theta]}}[\cdot]$ is the expectation with respect to the joint PDF $f(\boldsymbol{[r_0]},\boldsymbol{[\theta]})\!=\!f({r_0}_1,\cdots,{r_0}_N,\theta_1,\cdots,\theta_N)$. Since ${r_0}_j$'s and $\theta_j$'s are i.i.d, we have $f(\boldsymbol{[r_0]},\boldsymbol{[\theta]})=f_{\mathcal{R}_0}({r_0}_j)f_{\Theta}(\theta_j)\prod_{i\neq j}f_{\mathcal{R}_0}({r_0}_i)f_{\Theta}(\theta_i)$, where $f_{\mathcal{R}_0}({r_0}_j)$ and $f_{\Theta}(\theta_j)$ are described in Section~\ref{sec2}. The expectation can go inside the summation, then, we have  $\mathbb{E}_{\boldsymbol{[r_0]},\boldsymbol{[\theta]}}\left[\overline{R}_{{\rm{d}},j}(t_{\rm{u}}) \right]=\mathbb{E}_{{r_0}_j,\theta_j}\left[ \overline{R}_{{\rm{d}},j}(t_{\rm{u}})\right]$ for downlink and $\mathbb{E}_{\boldsymbol{[r_0]},\boldsymbol{[\theta]}}\left[\overline{R}_{{\rm{u}},j}(t_{\rm{u}}) \right]=\mathbb{E}_{{r_0}_j,\theta_j}\left[ \overline{R}_{{\rm{u}},j}(t_{\rm{u}})\right]$ for uplink. Since ${r_0}_j$'s and $\theta_j$'s are i.i.d, then:\vspace{-0.2cm}
\begin{equation*}
	\begin{aligned}
		&\mathbb{E}_{{r_0}_1,\theta_1}\!\!\left[ \overline{R}_{{\rm{d}},1}(t_{\rm{u}})\right]\!=\!\cdots\!=\!\mathbb{E}_{{r_0}_N,\theta_N}\!\left[ \overline{R}_{{\rm{d}},N}(t_{\rm{u}})\right]\! \triangleq\! \mathbb{E}_{{r_0},\theta}\! \left[ \overline{R}_{{\rm{d}}}(t_{\rm{u}})\right]\\\vspace{-0.2cm}
		&\mathbb{E}_{{r_0}_1,\theta_1}\!\!\left[ \overline{R}_{{\rm{u}},1}(t_{\rm{u}})\right]\!=\!\cdots\!=\!\mathbb{E}_{{r_0}_N,\theta_N}\!\left[ \overline{R}_{{\rm{u}},N}(t_{\rm{u}})\right]\! \triangleq\! \mathbb{E}_{{r_0},\theta}\! \left[ \overline{R}_{{\rm{u}}}(t_{\rm{u}})\right].\vspace{-0.1cm}
	\end{aligned}
\end{equation*} 
After substituting above equations in \eqref{equation17} and some manipulations, the OP can be expressed as:\vspace{-0.5cm}
\begin{equation}
	\label{equation17.1}
	\begin{aligned}
		&\max_{t_{\rm{u}}}\ \left( \overline{\mathcal{T}}\!= w_{\rm{u}}\mathbb{E}_{r_0,\theta}\left[ \overline{R}_{\rm{u}}(t_{\rm{u}})\right]\!+w_{\rm{d}}\mathbb{E}_{r_0,\theta}\left[\overline{R}_{\rm{d}}(t_{\rm{u}})\right]\right) ,\vspace{-0.1cm}
	\end{aligned}
\end{equation}\vspace{-0.2cm}
which is not dependent on any specific UEs.
The average is calculated over one update interval, since it is assumed the UE feeds back its velocity information to the AP after each update interval. The opposite behaviour of $\overline{R}_{\rm{u}}$ and $\overline{R}_{\rm{d}}$ with respect to the update interval (the former directly and the latter inversely are proportional to the update interval), results in an optimum point for $\overline{\mathcal{T}}$. In the following, $\overline{R}_{\rm{u}}$ and $\overline{R}_{\rm{d}}$ are calculated with some simplifying assumptions.

The exact and general state of SINR at the receiver is provided in \eqref{equation6}. However, for ease of analytical derivations, it can be simplified under some reasonable assumptions including: \textit{i}) the interference from other APs can be neglected due to employing FR plan, \textit{ii}) $H_{i,j,k}\approx H_{{\rm{LOS}},i,j}$. The latter assumption is based on the fact that in OWC systems, $H_{{\rm{LOS}},i,j}\!\! >>\!\! H_{{\rm{NLOS}},i,j}$. It was shown in Fig.~\ref{figu2} that the variation of the frequency response fluctuation around the LOS component is less than 1 dB. Using Fig.~\ref{fig1.2},
$\cos\phi_{ij}=\cos\psi_{ij}=h/d_{ij}$, can be substituted in \eqref{equation1}, then, the DC gain of the LOS channel is $H_{{\rm{LOS}},{i,j}}=G_0/d_{ij}^{m+3}$, where  $G_0\!=\!\frac{(m+1)A}{2\pi\sin^2\Psi_{\rm{c}}}g_{\rm{f}}\varsigma^2 h^{m+1}$. 
Hence, the approximate and concise equation of SINR at $k$th subcarrier of $j$th UE is given by:\vspace{0.1cm}
\begin{equation}
	\label{equation7}
	\vspace{-2pt}
	\gamma_{j,k}\approx\frac{Ge^{\frac{-4\pi kB_{{\rm{d}},n}}{\mathcal{K}w_0}}}{\left(r_j^2+h^2\right)^{m+3}},
\end{equation}
where $G=\frac{\mathcal{K}G_0^2R_{\rm{PD}}^2P_{\rm{d,opt}}^2}{(\mathcal{K}-2)\eta^2N_0B_{{\rm{d}},n}}$ and $r_j$ is the distance between the ${\rm{UE}}_j$ and the center of the cell which is located in it. Substituting \eqref{equation7} in \eqref{equation5}, the downlink throughput is given as:\vspace{0.1cm}
\begin{equation}
\label{equation8}
R_{{\rm{d}},j}=\frac{B_{{\rm{d}},n}}{\mathcal{K}}\!\sum_{k=1}^{\frac{\mathcal{K}}{2}-1}\!\log_2\!\left(\! 1+s_{j,k}\frac{Ge^{\frac{-4\pi kB_{{\rm{d}},n}}{\mathcal{K}w_0}}}{\left(r_j^2+h^2\right)^{m+3}}\right) .
\end{equation}
Noting that typically in LiFi cellular networks using FR, SINR values are high enough, we have:\vspace{-0.2cm}
\begin{equation}
\label{equation9}
R_{{\rm{d}},j}=\frac{B_{{\rm{d}},n}}{\mathcal{K}}\!\sum_{k=1}^{\frac{\mathcal{K}}{2}-1}\!\!s_{j,k}\log_2\left(\frac{Ge^{\frac{-4\pi kB_{{\rm{d}},n}}{\mathcal{K}w_0}}}{\left(r_j^2+h^2\right)^{m+3}}\right) .
\end{equation}
	
Same approximations can be also considered for uplink throughput. Define $G_{\rm{u}}\!=\!\frac{(G_0R_{\rm{PD}}P_{\rm{u,opt}})^2}{\eta^2N_0B_{{\rm{u}},n}}$, then, the SINR at the AP is $\gamma_{{\rm{u}},j}\!=\!{G_{\rm{u}}}/{(r_j^2+h^2)^{m+3}}$. Substituting it in \eqref{equation16}, the uplink throughput of ${\rm{UE}}_j$ can approximately be obtained as:\vspace{-0.1cm}
\begin{equation}
\label{equation12}
R_{{\rm{u}},j}\cong \left( 1-\dfrac{t_{\rm{fb}}}{t_{\rm{u}}}\right)  \frac{\widetilde{\mathcal{T}}_{\rm{u}}B_{{\rm{u}},n}}{N}\log_2\!\left(\!\frac{G_{\rm{u}}}{\left(r_j^2+\!h^2\right)^{m+3}}\! \!\right).\vspace{-0.1cm}
\end{equation}
	
Without loss of generality and for ease of notations, we consider one of the $N$ UEs for the rest of derivations and remove the subscript $j$. 
The average uplink throughput over one update interval is given as:\vspace{-0.1cm}
\begin{equation}
	\label{equation18}
	\begin{aligned}
		&\overline{R}_{\rm{u}}\!=\!\left(\!1\!-\dfrac{t_{\rm{fb}}}{t_{\rm{u}}} \right)\!\frac{\widetilde{\mathcal{T}}_{\rm{u}}B_{{\rm{u}},n}}{N}\dfrac{1}{t_{\rm{u}}}\int_{0}^{t_{\rm{u}}}\!\!\!\!\log_2\!\left(\frac{G_{\rm{u}}}{(r^2(t)+h^2)^{m+3}}\right){\rm{d}}t\\
		&=\!\dfrac{2(m+3)\widetilde{\mathcal{T}}_{\rm{u}}B_{{\rm{u}},n}}{N}\!\left(\!1\!-\dfrac{t_{\rm{fb}}}{t_{\rm{u}}}\! \right) \!\!\left(\! \dfrac{1}{2(m+3)} \log_2\!\left(\!\dfrac{G_{\rm{u}}}{(r^2(t_{\rm{u}})+h^2)^{m+3}} \right)+\dfrac{r_0\cos\theta}{2vt_{\rm{u}}}\!\log_2\!\! \left(\!\dfrac{r^2(t_{\rm{u}})\!+\!h^2}{r_0^2+h^2}\! \right)\!\right.\\ 
		&\left.+\frac{1} {\ln(2)}-\frac{\!(h^2\!+r_0^2\sin^2\theta)^{\frac{1}{2}}}{vt_{\rm{u}}\ln(2)}\tan^{-1}\!\!\left(\!\! \frac{vt_{\rm{u}}-r_0\cos\theta}{(h^2\!+r_0^2\sin^2\theta)^{\frac{1}{2}}}\!\!\right) \! -\frac{\!(h^2\!+\!r_0^2\sin^2\!\theta)^{\frac{1}{2}}}{vt_{\rm{u}}\ln(2)}\!\tan^{-1}\!\!\left(\!\!\frac{r_0\cos\theta}{(h^2\!+\!r_0^2\sin^2\!\theta)^{\frac{1}{2}}} \!\!\right)\!\! \right). \vspace{-0.2cm}	
	\end{aligned} 
\end{equation}
	
The average downlink throughput over one update interval can be obtained as: \vspace{-0.1cm}
\begin{equation}
	\label{equation19}
	\begin{aligned}
		&\overline{R}_{\rm{d}}=\frac{B_{{\rm{d}},n}}{\mathcal{K}t_{\rm{u}}}\!\int_{0}^{t_{\rm{u}}}\sum_{k=1}^{k_{\rm{req}}}\!\log_2\left(\dfrac{Ge^{\frac{-4\pi kB_{{\rm{d}},n}}{\mathcal{K}w_0}}}{\left(r^2(t)+h^2\right)^{m+3}}\right) {\rm{d}}t
		=\frac{k_{\rm{req}}B_{{\rm{d}},n}}{\mathcal{K}t_{\rm{u}}}\!\int_{0}^{t_{\rm{u}}}\log_2\left( \frac{Ge^{{-2\pi(k_{\rm{req}}+1) B_{{\rm{d}},n}}/{\mathcal{K}w_0}}} {(r^2(t)+h^2)^{m+3}} \right){\rm{d}}t\\
		&\resizebox{1\hsize}{!}{$=\!\dfrac{2(m+3)k_{\rm{req}}B_{{\rm{d}},n}}{\mathcal{K}}\!\left(\!\dfrac{1}{2(m+3)} \log_2\!\left(\!\dfrac{Ge^{{-2\pi(k_{\rm{req}}+1) B_{{\rm{d}},n}}/{\mathcal{K}w_0}}}{(r^2(t_{\rm{u}})+h^2)^{m+3}} \right)\!+\dfrac{\!r_0\cos\theta}{2vt_{\rm{u}}}\!\log_2\!\! \left(\!\dfrac{r^2(t_{\rm{u}})\!+\!h^2}{r_0^2+h^2}\! \right)\!+\dfrac{1} {\ln(2)}\right.$}\\ 
		&\left.-\dfrac{(h^2+r_0^2\sin^2\theta)^{\frac{1}{2}}}{vt_{\rm{u}}\ln(2)}\tan^{-1}\left( \dfrac{vt_{\rm{u}}-r_0\cos\theta}{(h^2+r_0^2\sin^2\theta)^{\frac{1}{2}}}\right) -\dfrac{\!(h^2\!+\!r_0^2\sin^2\!\theta)^{\frac{1}{2}}}{vt_{\rm{u}}\ln(2)}\!\tan^{-1}\!\!\left(\!\!\dfrac{r_0\cos\theta}{(h^2\!+\!r_0^2\sin^2\!\theta)^{\frac{1}{2}}} \!\!\right)\!\! \right). \vspace{-0.2cm}
  \end{aligned}
\end{equation}
where $k_{\rm{req}}$ is the required number of subcarriers to be allocated to the UE at $t=0$. With the initial and random distance of $r_0$ from the cell center, the required number of subcarriers can approximately be obtained as:\vspace{-0.25cm}
\begin{equation}
\label{equation20}
k_{\rm{req}}\cong \frac{\mathcal{K}R_{\rm{req}}}{B_{{\rm{d}},n}\log_2\left(G/(r_0^2+h^2)^{m+3} \right) }
\end{equation}
The exact value and proof are given in Appendix\ref{App-A}. Both the average uplink and downlink throughput given in \eqref{equation18} and \eqref{equation19}, respectively, are continuous and derivative in the range $(0,2r_{\rm{c}}/v)$. Therefore, we can express the following proposition to find the optimal update interval that results in the maximum sum-throughput.
	
\textbf{Proposition.} \textit{Let $t_{\rm{u}}$ be continuous in the range of $(0,2r_{\rm{c}}/v)$. The optimal solution to the OP given in \eqref{equation17.1} can be obtained by solving the following equation:
	\begin{equation}
	\label{equation21}
	\mathbb{E}_{r_0,\theta}\left[ \frac{\partial \overline{\mathcal{T}}}{\partial t_{\rm{u}}}\right] =w_{\rm{u}}\mathbb{E}_{r_0,\theta}\left[ \frac{\partial \overline{\mathcal{T}}_{\rm{u}}}{\partial t_{\rm{u}}}\right] +w_{\rm{d}}\mathbb{E}_{r_0,\theta}\left[ \frac{\partial \overline{\mathcal{T}}_{\rm{d}}}{\partial t_{\rm{u}}}\right] =0.
	\end{equation}  
	For $vt_{\rm{u}}\ll h$, the root of \eqref{equation21} can be well approximated as:
	\begin{equation}
	\label{equation22}
	\begin{aligned}
	\!\!\!\widetilde{t}_{\rm{u,opt}}\!\cong\!\! \left(\!\dfrac{\frac{\!3\!\ln(2)}{2(m+3)}w_{\rm{u}}t_{\rm{fb}}\widetilde{\mathcal{T}}_{\rm{u}}B_{{\rm{u}},n}C_1 }{w_{\rm{d}}v^2NR_{\rm{req}}+C_2w_{\rm{u}}v^2\widetilde{\mathcal{T}}_{\rm{u}}B_{{\rm{u}},n}}\!\! \right)^{\!\!\frac{1}{3}},
	\end{aligned}
	\end{equation}
	where
	\begin{equation}
	\label{equation23}
	\begin{aligned}
	&C_1=\dfrac{\mathbb{E}_{r_0}\!\!\left[\log_2\!\!\left(\!\dfrac{G_{\rm{u}}}{(r_0^2+h^2)^{m+3}}\!\right)\right]\mathbb{E}_{r_0}\!\!\left[ \log_2\!\!\left(\!\dfrac{G}{(r_0^2+h^2)^{m+3}}\!\right)\!\right]}{\mathbb{E}_{r_0,\theta}\!\!\left[\dfrac{(h^2+r_0^2\sin^2\theta)^2}{(h^2+r_0^2)^3}\right] }, \ \
	C_2=\mathbb{E}_{r_0}\!\!\left[ \log_2\!\!\left(\!\dfrac{G}{(r_0^2+h^2)^{m+3}}\!\right)\!\right].
	\end{aligned}
	\end{equation}	
}
\textit{Proof: See Appendix\ref{App-B}}

	\begin{table}[t!]
		\centering
		\caption{Uplink simulation parameters}
		\vspace{-4pt}
		\label{TableII}
		\vspace{-2pt}
		{\raggedright
			\vspace{4pt} \noindent
			\begin{tabular}{p{150pt}|p{25pt}|p{35pt}}
				\hline
				\parbox{150pt}{\centering{\small Parameter}} & \parbox{25pt}{\centering{\small Symbol}} & \parbox{35pt}{\centering{\small Value}} \\
				\hline
				\hline
				\parbox{150pt}{\raggedright{\small Transmitted uplink optical power}} & \parbox{25pt}{\centering{\small $P_{\rm{u,opt}}$}} & \parbox{35pt}{\centering{\small $0.2$ W}} \\
				\hline
				\parbox{125pt}{\raggedright{\small Uplink FR bandwidth}} & \parbox{25pt}{\centering{\small $B_{{\rm{u}},n}$}} & \parbox{35pt}{\centering{\small $5$ MHz}} \\
				\hline
				\parbox{150pt}{\raggedright{\small Average length of uplink payload}} & \parbox{25pt}{\centering{\small $L_{\rm{D}}$}} & \parbox{35pt}{\centering{\small $2000$ B}} \\
				\hline
				\parbox{125pt}{\raggedright{\small Physical header}} & \parbox{25pt}{\centering{\small $H_{\rm{PHY}}$}} & \parbox{35pt}{\centering{\small $128$ b}} \\
				\hline
				\parbox{125pt}{\raggedright{\small MAC header}} & \parbox{25pt}{\centering{\small $H_{\rm{MAC}}$}} & \parbox{35pt}{\centering{\small $272$ b}} \\
				\hline
				\parbox{125pt}{\raggedright{\small RTS packet size}} & \parbox{25pt}{\centering{\small $L_{\rm{RTS}}$}} & \parbox{35pt}{\centering{\small $288$ b}} \\
				\hline
				\parbox{125pt}{\raggedright{\small CTS packet size}} & \parbox{25pt}{\centering{\small $L_{\rm{CTS}}$}} & \parbox{35pt}{\centering{\small $240$ b}} \\
				\hline
				\parbox{125pt}{\raggedright{\small ACK packet size}} & \parbox{25pt}{\centering{\small $L_{\rm{ACK}}$}} & \parbox{35pt}{\centering{\small $240$ b}} \\
				\hline
				\parbox{125pt}{\raggedright{\small SIFS}} & \parbox{25pt}{\centering{\small $--$}} & \parbox{35pt}{\centering{\small $16\ \mu$s}} \\
				\hline
				\parbox{125pt}{\raggedright{\small DIFS}} & \parbox{25pt}{\centering{\small $--$}} & \parbox{35pt}{\centering{\small $32\ \mu$s}} \\
				\hline
				\parbox{125pt}{\raggedright{\small Backoff slot duration}} & \parbox{25pt}{\centering{\small $t_{\rm{slot}}$}} & \parbox{35pt}{\centering{\small $8\ \mu$s }} \\
				\hline
				\parbox{125pt}{\raggedright{\small Propagation delay}} & \parbox{25pt}{\centering{\small $t_{\rm{delay}}$}} & \parbox{35pt}{\centering{\small $1\ \mu$s }} \\
				\hline
				\parbox{125pt}{\raggedright{\small Feedback time}} & \parbox{25pt}{\centering{\small $t_{\rm{fb}}$}} & \parbox{35pt}{\centering{\small $0.8$ ms }} \\
				\hline
			\end{tabular}
			\vspace{-2pt}
		}
		\vspace{-1cm}
	\end{table}
	
\begin{figure}[t!]
		\centering
		\subfloat[The effect of UE's velocity on optimal update interval $(P_{\rm{d,opt}}=8 \rm{watt})$. \label{sub1:13a}]{%
			\includegraphics[width=75mm,height=35mm]{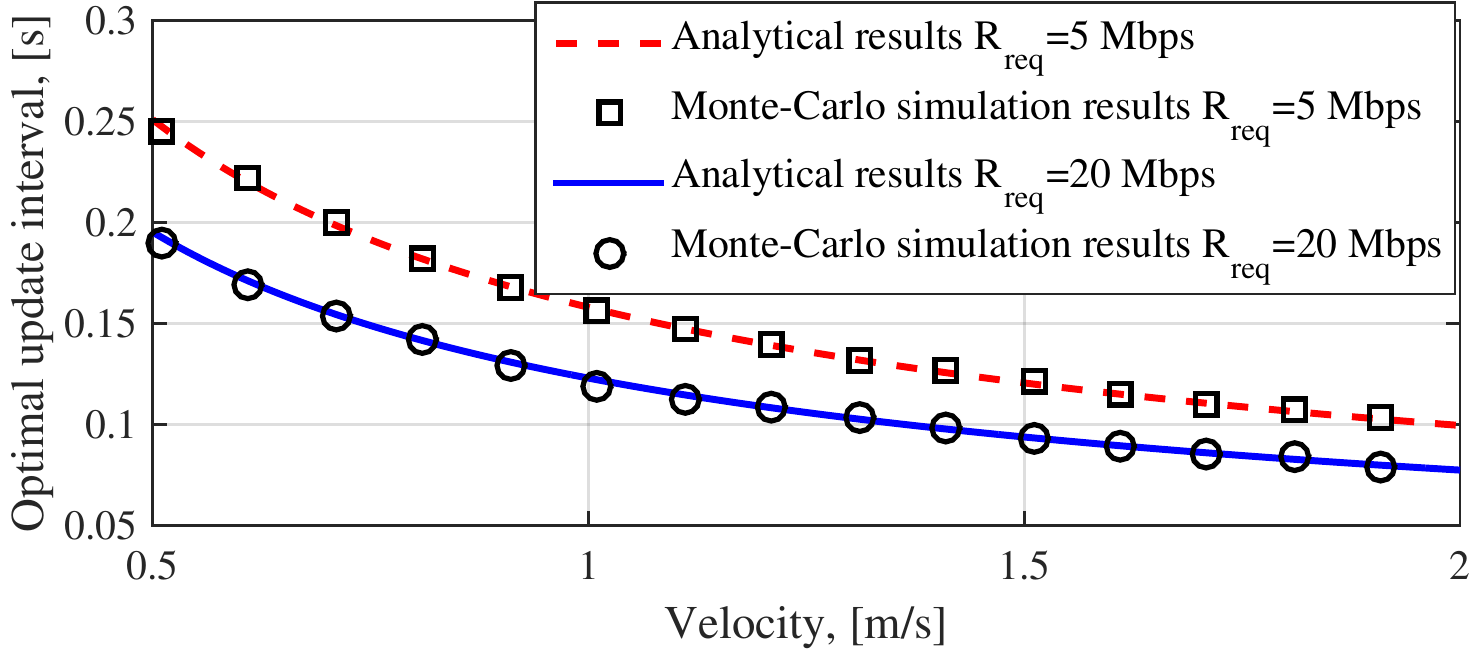}
		}\ \
		\subfloat[The effect of transmitted downlink optical power on optimal update interval $(v=1$ m/s$)$. \label{sub1:13b}]{%
			\includegraphics[width=75mm,height=35mm]{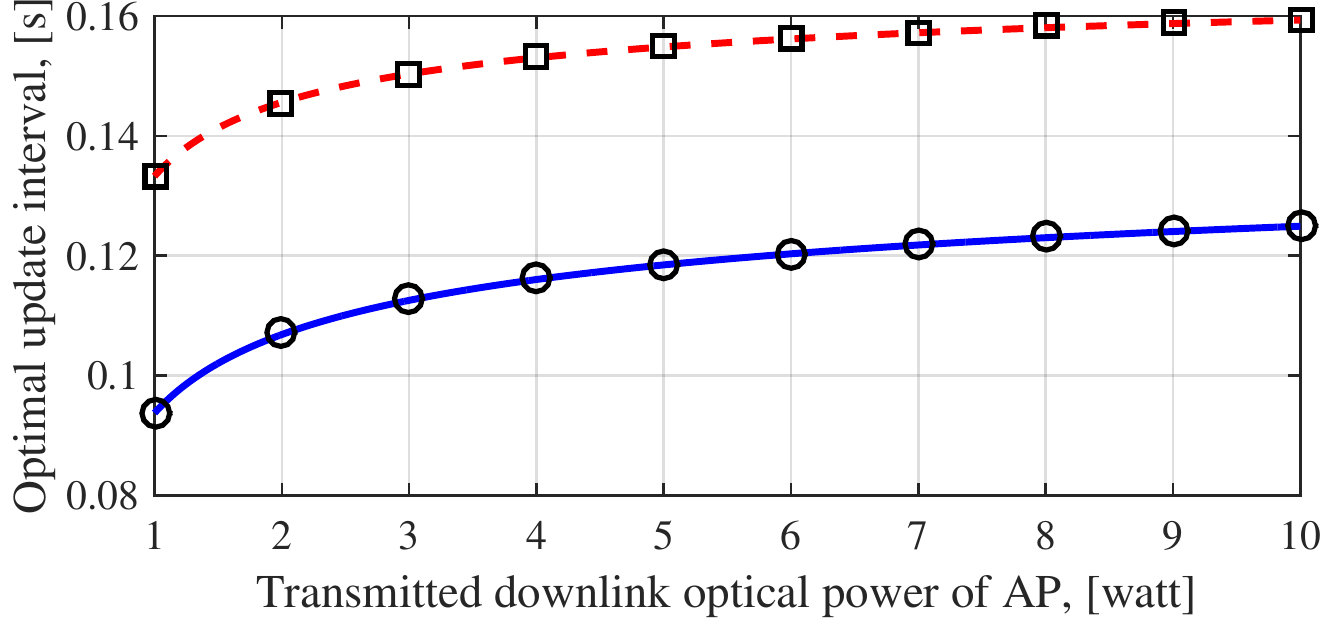}
		}\\ \vspace{-0.2cm}
		\caption{The effects of UE's velocity and downlink optical power on optimal update interval for $R_{\rm{req}}=5$ Mbps and $R_{\rm{req}}=20$ Mbps, and $N\!=\!5$.}
		\label{fig13}
		\vspace{-1.4cm}
	\end{figure}

As it can be seen from \eqref{equation22}, the optimum update interval depends on both physical and MAC layer parameters. Among them, the UE velocity affects the update interval more than others. Let's fix the other parameters, then $\widetilde{t}_{\rm{u,opt}}={C_{\rm{const}}}/{v^{\frac{2}{3}}}$, where 
$C_{\rm{const}}\!=\left(\!\frac{\!\frac{3\!\ln(2)}{2(m+3)}w_{\rm{u}}t_{\rm{fb}}\widetilde{\mathcal{T}}_{\rm{u}}B_{{\rm{u}},n}C_1 }{w_{\rm{d}}R_{\rm{req}}+C_2w_{\rm{u}}\widetilde{\mathcal{T}}_{\rm{u}}B_{{\rm{u}},n}} \right)^{\!\!\frac{1}{3}}$. We study the effect of UE's velocity and transmitted downlink optical power on the update interval as illustrated in Fig.~\ref{fig13}. Analytical and Monte-Carlo simulations are presented for $w_{\rm{u}}=w_{\rm{d}}$, $N\!=\!5$ and with the downlink and uplink simulation parameters given in Table~\ref{TableI} and Table~\ref{TableII}, respectively. For a fixed $t_{\rm{u}}$, Monte-Carlo simulations are accomplished $10^4$ times, where in each run, the UE's initial position and direction of movement are randomly chosen. Then, for the considered $t_{\rm{u}}$, the expected sum-throughput, $\overline{\mathcal{T}}$, can be obtained by averaging out over $10^4$ runs. Afterwards, based on the greedy search and for different  $t_{\rm{u}}$, varying in the range $0\!<\!t_{\rm{u}}\!<\!2r_{\rm{c}}/v$, Monte-Carlo simulations are repeated. The optimal update interval corresponds to the maximum sum-throughput. The effect of UE's velocity on optimal update interval for $R_{\rm{req}}\!=\!5$ Mbps and $R_{\rm{req}}\!=\!20$ Mbps is shown in Fig.~\ref{fig13}-(a). Here, we can see the optimal update interval decrease rapidly as UE's speed increases, according to $v^{-2/3}$. Further, Monte-Carlo simulations confirm the accuracy of analytical results provided in \eqref{equation22}.
Fig.~\ref{fig13}-(b) illustrates the saturated effect of transmitted optical power on $\widetilde{t}_{\rm{u,opt}}$. As can be observed, the variation of optimal update interval due to alteration of $P_{\rm{d,opt}}$ is less than $30$ ms. From both Fig.~\ref{fig13}-(a) and Fig.~\ref{fig13}-(b), it can be deduced the lower $R_{\rm{req}}$, the higher $\widetilde{t}_{\rm{u,opt}}$.\vspace{-0.0cm}
	
Now let's consider an overloaded multi-user scenario with $N$ users. The fair scheduler introduced in \eqref{equation4} tries to equalize the rate of all UEs. For high number of subcarriers, the UEs achieve approximately the same data rate.   Accordingly, the on average achieved data rate of UEs in an overloaded network for high number of subcarriers would nearly be $\lambda R_{\rm{req}}$, where $0\!<\!\lambda\!<\!1$.
This system is equivalent to a non-overloaded multi-user system where all UEs have achieved on average their request rate of $\lambda R_{\rm{req}}$. Then, the approximate optimal update interval that results in near-maximum sum-throughput is given as:\vspace{-0.1cm}
	\begin{equation}
		\label{equation25}
		\begin{aligned}
			\!\!\!\widetilde{t}_{\rm{u,opt}}\!\cong\!\! \left(\!\dfrac{\frac{\!3\!\ln(2)}{2(m+3)}w_{\rm{u}}t_{\rm{fb}}\widetilde{\mathcal{T}}_{\rm{u}}B_{{\rm{u}},n}C_1 }{w_{\rm{d}}v^2 N\lambda R_{\rm{req}} +C_2w_{\rm{u}}v^2\widetilde{\mathcal{T}}_{\rm{u}}B_{{\rm{u}},n}}\!\! \right)^{\!\!\frac{1}{3}}.
		\end{aligned}
		\vspace{-0.2cm}
	\end{equation}
	\vspace{-0.38cm}
	
	\begin{figure}[t!]
		\centering
		\includegraphics[width=85mm,height=50mm]{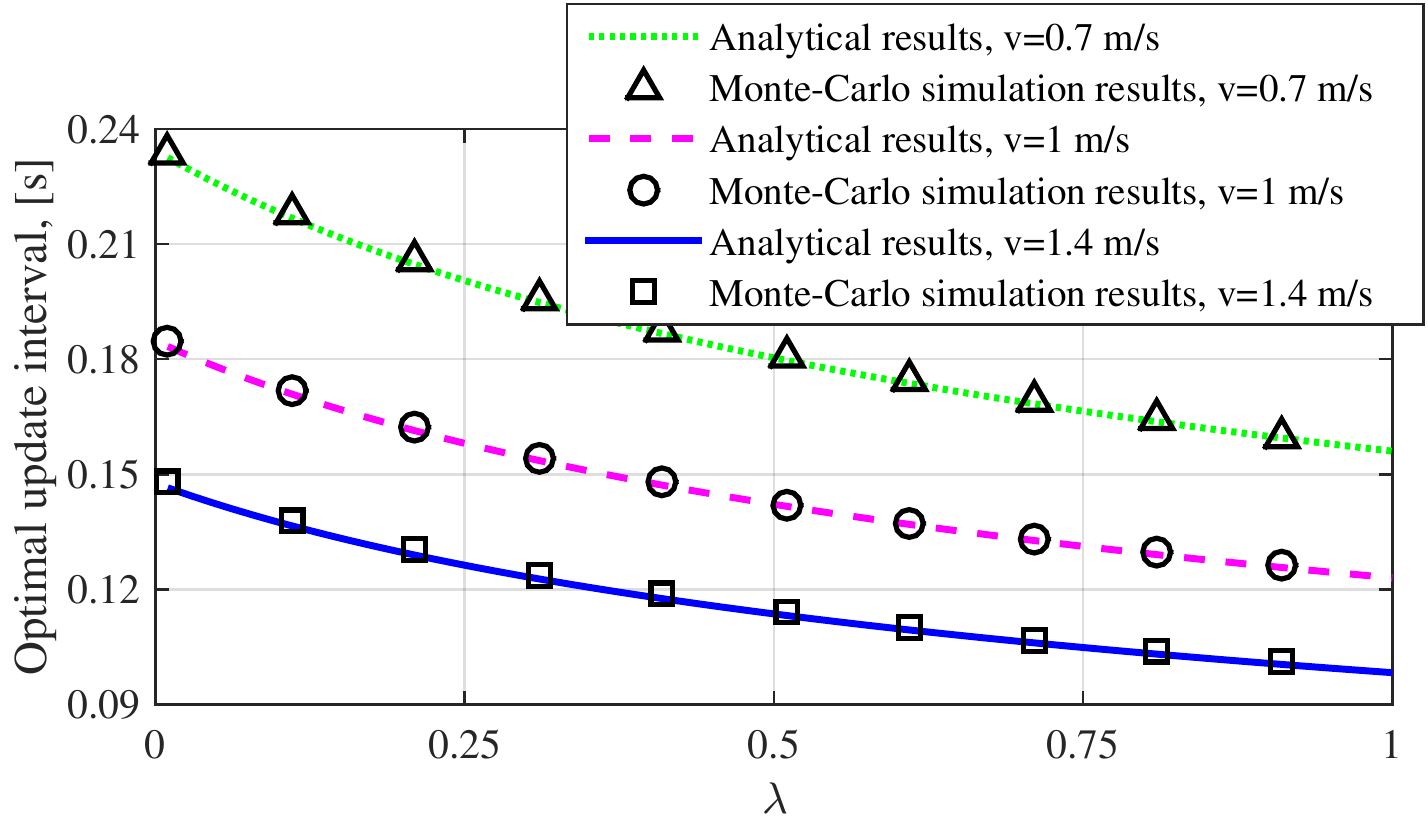}\vspace{-.25cm}
		\caption{Optimal update interval versus overload parameter, $\lambda$, for different UE's velocity $(N=5)$. } 
		\label{fig8}
		\vspace{-1cm}  
	\end{figure}
	
Analytical and Monte-Carlo simulations of an overloaded system are shown in Fig.~\ref{fig8}. Three speed values are chosen around the average human walking speed which is $1.4$ m/s \cite{bohannon1997comfortable}. Note that to obtain an overloaded system either the number of UEs or their request data rate can be increased. In the results shown in Fig.~\ref{fig8}, we fixed the number of UEs to $N=5$ and increase their $R_{\rm{req}}$. As can be inferred from these results, as the network becomes more overloaded, the optimal update interval should be increased. The reason is that in an overloaded network, due to lack of enough resources updating the AP frequently is useless and it just wastes the uplink resources. 
	
	\begin{figure}[t!]
		\centering
		\resizebox{0.45\linewidth}{!}{
			\includegraphics{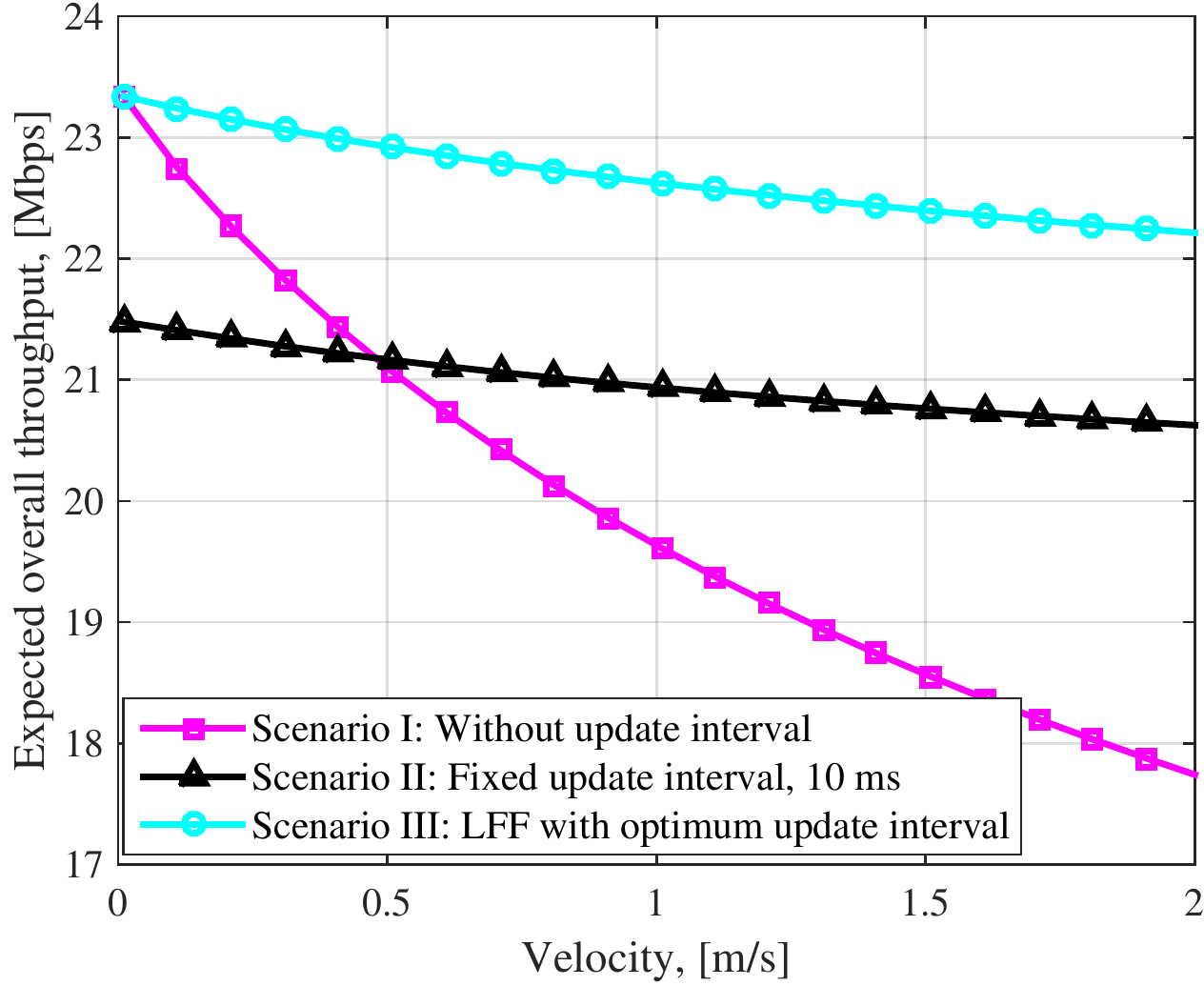}}\vspace{-.25cm}
		\caption{Expected overall throughput versus UE's velocity for three scenarios; and $R_{\rm{req}}=20$ Mbps, $w_{\rm{u}}=w_{\rm{d}}=1$. } 
		\label{fig12.2}
		\vspace{-1.3cm}  
	\end{figure}
	
To verify the significance of update interval in practical systems, three scenarios have been considered. Scenario I: a system without any update interval; 
Scenario II: a system with the conventional fixed update interval but without looking at the UE's velocity; 
Scenario III: a system with the proposed update interval and adjustable with the UE's velocity. For these scenarios, Monte-Carlo simulation results of expected sum-throughput versus different UE's velocity have been obtained and presented in Fig.~\ref{fig12.2}. In scenario I, the UEs only update the AP once at the start of the connection by transmission of the SINR information of $\mathcal{K}/2-1$ subcarriers. 
For scenario II, the fixed update interval is considered to be $t_{\rm{u}}=10$ ms and independent of UE's velocity. Fixed update interval is currently used in LTE with $t_{\rm{u}}=10$ ms by transmission of one-bit feedback information at the beginning of every frame \cite{3gpp.36.213}. It is worth mentioning that for practical wireless systems, it is common to transmit feedback frequently, e.g., at the beginning of each frame regardless of the UE channel variation and velocity. As can be seen from the results, the proposed LFF scheme outperforms the conventional method with fixed update interval. For low speeds (up to $0.5$ m/s), the conventional fixed update interval even falls behind the system without any update interval. This is due to redundant feedback information being sent to the AP. The gap between LFF and scenario II with fixed update interval is due to both higher uplink and downlink throughput of LFF. LFF provides higher uplink throughput thanks to transmission of lower feedback compared to fixed update interval scheme. Also, in scenario II, the UEs after $10$ ms update the AP with one bit per subcarrier, and the AP does not know the SINR value of each subcarrier to allocate them efficiently to the UEs.   \vspace{-0.2cm}

\subsection{LF Schemes Comparison}
\begin{figure}[t!]
		\centering
		\resizebox{0.5\linewidth}{!}{
			\includegraphics{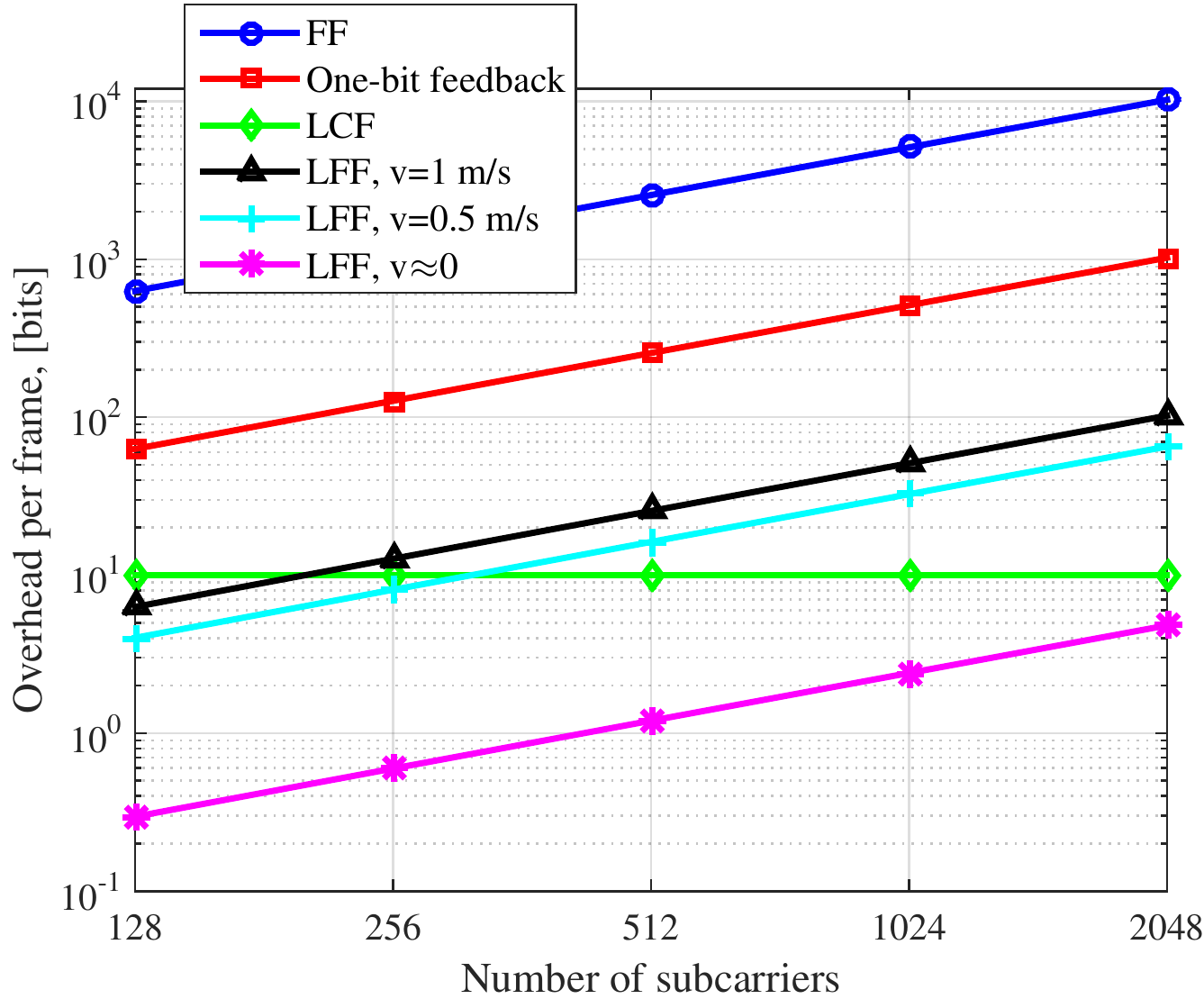}}\vspace{-0.2cm}
		\caption{Transmitted overhead versus different number of subcarriers.} 
		\label{fig5}
		\vspace{-1cm}  
	\end{figure}

A comparison between the FF, one-bit, LCF and LFF schemes in case of transmitted overhead is given in Table~\ref{TableIII}. It is assumed that the SINR on each subcarrier can be fedback to the AP using $\mathscr{B}$ bits, and $M=[\widetilde{t}_{\rm{u,opt}}/t_{\rm{fr}}]$. Note that for $M\geq (\mathscr{B}+1)$, the overhead per frame of the LFF scheme is lower than the one-bit feedback technique. Also, for $M\geq \mathcal{K}/2$, LFF scheme produces lower overhead per frame in comparison to LCF. For $N=5$, $\mathscr{B}=10$, $t_{\rm{fr}}=1.6$ ms and $R_{\rm{req}}=5$ Mbps the overhead per frame versus different number of subcarriers are illustrated in Fig.~\ref{fig5}. The rest of parameters are the same as given in Table~\ref{TableI} and Table~\ref{TableII}.
As can be observed from Fig.~\ref{fig5}, the FF scheme generates huge amount of feedback overhead especially for high number of subcarriers. The practical one-bit feedback reduces the overhead by a factor of $\mathscr{B}$. As can be seen, the LCF always falls below the one-bit feedback method. The gap between LCF and one-bit feedback becomes remarkable for higher number of subcarriers. The overhead results of the LFF have been also presented for stationary UEs and UEs with low and normal speed. Clearly, the LFF generates the lower feedback overhead per frame as the UE's velocity tends to zero. 
The expected sum-throughput of different feedback schemes with the same parameters as mentioned above are summarized in Table.~\ref{TableIV}. As we expected, the LFF outperforms the other schemes when the UEs are stationary. However, the sum-throughput of the LCF method is higher for mobile UEs.  \vspace{-0.2cm}
	\begin{table}[t!]
		\centering
		\caption{Comparison of feedback schemes in case of overhead}
		\vspace{-10pt}
		\label{TableIII}
		\vspace{-2pt}
		{\raggedright
			\vspace{4pt} \noindent
			\begin{tabular}{p{40pt}||p{80pt}|p{80pt}|p{80pt}|p{100pt}}
				\hline
				\parbox{40pt}{\centering{\small Scheme}} & 
				\parbox{80pt}{\centering{\small Full feedback}}&
				\parbox{80pt}{\centering{\small One-bit feedback}}&
				\parbox{80pt}{\centering{\small Proposed LCF}}&
				\parbox{100pt}{\centering{\small Proposed LFF}}\\				
				\hline
				\hline
				\parbox{40pt}{\centering{\small Overhead}} & 
				\parbox{80pt}{\centering{\small $\mathscr{B}(\mathcal{K}/2-1)$ bpf}}& 
				\parbox{80pt}{\centering{\small $(\mathcal{K}/2-1)$ bpf }}& 
				\parbox{80pt}{\centering{\small $\mathscr{B}$ bpf }}& 
				\parbox{100pt}{\centering{\small $\mathscr{B}(\mathcal{K}/2-1)/M$ bpf }}\\
				\hline
			\end{tabular}
			\vspace{2pt}
		}
		\vspace{-0.5cm}
	\end{table}
	
		\begin{table}[t!]
			\centering
			\caption{Comparison of feedback schemes in case of expected sum-throughput, $N=5$, $R_{\rm{req}}=5$ Mbps and $w_{\rm{u}}=w_{\rm{d}}=1$.}
			\vspace{-10pt}
			\label{TableIV}
			\vspace{-2pt}
			{\raggedright
				\vspace{4pt} \noindent
				\begin{tabular}{p{160pt}||p{60pt}|p{70pt}|p{60pt}|p{60pt}}
					\hline
					\parbox{160pt}{\centering{\small Scheme}} & 
					\parbox{60pt}{\raggedright{\small Full feedback}} &
					\parbox{70pt}{\raggedright{\small One-bit feedback}} &
					\parbox{60pt}{\raggedright{\small Proposed LCF}} & 
					\parbox{60pt}{\raggedright{\small Proposed LFF}} \\
					\hline
					\hline
					\parbox{160pt}{\centering{\small Expected sum-throughput ($v=0$ m/s)}}&
					\parbox{60pt}{\centering{\small $6.67$ Mbps }} &
					\parbox{70pt}{\centering{\small $7.64$ Mbps }} &
					\parbox{60pt}{\centering{\small $8.33$ Mbps }} &
					\parbox{60pt}{\centering{\small $8.35$ Mbps }} \\
					\hline
					\parbox{160pt}{\centering{\small Expected sum-throughput ($v=1$ m/s)}}&
					\parbox{60pt}{\centering{\small $6.67$ Mbps }}&
					\parbox{70pt}{\centering{\small $7.47$ Mbps }}&
					\parbox{60pt}{\centering{\small $8.33$ Mbps }}&
					\parbox{60pt}{\centering{\small $8.08$ Mbps }}\\
					\hline
				\end{tabular}
				\vspace{2pt}
			}
			\vspace{-0.0cm}
		\end{table}

\section{Conclusion and Future Works}
\vspace{-0.1cm}
\label{sec9}
%For optimum resource allocation and scheduling continuous feedback from all UEs is required at the AP. 
Two methods for reducing feedback cost were proposed in this paper: \textit{i})
the limited-content feedback (LCF) scheme, and \textit{ii}) the limited-frequency feedback (LFF) method. The former is based on reducing the content of feedback information by only sending the SINR of the first subcarrier and estimating the SINR of other subcarriers at the AP. The latter is based on the less frequent transmission of feedback information. The optimal update interval was derived, which results in maximum expected sum-throughput of uplink and downlink. The Monte-Carlo simulations confirmed the accuracy of analytical results. The effect of different parameters on optimum update interval was studied.
It was also shown that the proposed LCF and LFF schemes provide better sum-throughput while transmitting lower amount of feedback compared to the practical one-bit feedback method. The combination of the LCF with the update interval is the  topic of our future study.

\appendices
\section*{Appendix}
\subsection{Proof of \eqref{equation20}}
\label{App-A} 
According to the RWP mobility model, the UE is initially located at ${\bf{P}}_{0}$ with the distance $r_0$ from cell center. The scheduler at the AP is supposed to allocate the resources to the UEs as much as they require. Thus, the achievable data throughput of the UE at $t=0$  is equal to the requested data rate i.e., $R(0)=R_{\rm{req}}$. Hence, $k_{\rm{req}}$ can be obtained by solving the following equation:
\vspace{-1pt}
	\begin{equation}
		\label{equation29}
		\begin{aligned}
			&R_{\rm{req}}=\frac{B_{{\rm{d}},n}}{\mathcal{K}}\sum_{k=1}^{k_{\rm{req}}}\log_2\!\!\left(\!\frac{Ge^{\frac{-4\pi kB_{{\rm{d}},n}}{\mathcal{K}w_0}}} {(h^2+r_0^2)^{m+3}}\right) 
			=\!\frac{B_{{\rm{d}},n}}{\mathcal{K}}\!\sum_{k=1}^{k_{\rm{req}}}\!\log_2\!\!\left(\!\frac{G}{(h^2+r_0^2)^{m+3}}\!\right)\!+\!\frac{B_{{\rm{d}},n}}{\mathcal{K}}\!\sum_{k=1}^{k_{\rm{req}}}\!\log_2\!\left(\!e^{\frac{-4\pi kB_{{\rm{d}},n}}{\mathcal{K}w_0}}\!\right) \\
			&=\!\frac{k_{\rm{req}}B_{{\rm{d}},n}}{\mathcal{K}}\log_2\!\left(\!\frac{G}{(h^2+r_0^2)^{m+3}}\!\right)\!-\!\frac{4\pi }{w_0}\!\left(\!\frac{B_{{\rm{d}},n}}{\mathcal{K}}\! \right)^{\!\!2} \log_2\!e \sum_{k=1}^{k_{\rm{req}}}k\\
			&=\!\dfrac{k_{\rm{req}}B_{{\rm{d}},n}}{\mathcal{K}}\log_2\!\left(\!\dfrac{G}{(h^2+r_0^2)^{m+3}}\!\right)\!-\!\dfrac{2\pi }{w_0}\!\left(\!\dfrac{B_{{\rm{d}},n}}{\mathcal{K}}\! \right)^{\!\!2}\! (\log_2\!e) k_{\rm{req}}(k_{\rm{req}}+1)\\
			&\Rightarrow k_{\rm{req}}^2\!+\!\left(\!\!1\!-\frac{\log_2\!\left(\!\frac{G}{(h^2+r_0^2)^{m+3}}\right)}{\frac{2\pi B_{{\rm{d}},n} }{\mathcal{K}w_0}\log_2\!e }\! \right)\!\! k_{\rm{req}}\!+\!\frac{R_{\rm{req}}}{\frac{2\pi }{w_0}\!\left(\!\frac{B_{{\rm{d}},n}}{\mathcal{K}}\! \right)^{\!\!2}\! \log_2\!e}\!=\!0 .
		\end{aligned}
	\end{equation}
	
The above equation is a quadratic equation and it has two roots where the acceptable $k_{\rm{req}}$ can be obtained as follows:
	\vspace{-1pt}
	\begin{equation}
		\label{equation30}
		\begin{aligned}
			k_{\rm{req}}=
			\frac{\!\!\left(\!\frac{\log_2\!\left(\!\frac{G}{(h^2+r_0^2)^{m+3}}\right)}{\frac{2\pi B_{{\rm{d}},n} }{\mathcal{K}w_0}\log_2\!e }\!-\!1\! \right) \!\!-\!\!\sqrt{\!\!\left(\!\!1\!-\frac{\log_2\!\left(\!\frac{G}{(h^2+r_0^2)^{m+3}}\right)}{\frac{2\pi B_{{\rm{d}},n} }{\mathcal{K}w_0}\log_2\!e } \right)^2\!\!\!\!-\frac{4R_{\rm{req}}}{\frac{2\pi }{w_0}\!\left(\frac{B_{{\rm{d}},n}}{\mathcal{K}} \right)^{\!2}\!\! \log_2\!e}}}{2}.
		\end{aligned}
	\end{equation}
	
If $R_{\rm{req}}\ll\frac{w_0}{8\pi}\log_2\!\left(\!\frac{G}{(h^2+r_0^2)^{m+3}}\right)$, the approximate number of required subcarriers is $k_{\rm{req}}\cong \frac{\mathcal{K}R_{\rm{req}}}{B_{{\rm{d}},n}\log_2\left(\frac{G}{(h^2+r_0^2)^{m+3}}\right)}$. With the parameters given in Table~\ref{TableI}, the constraint on the requested data rate is $R_{\rm{req}}<\!<350$ Mbps.
\vspace{-0.3cm}
	
\subsection{Proof of Proposition}
\label{App-B} 
The optimal solution of the OP given in \eqref{equation17.1} can be obtained by finding the roots of its derivation that is $\frac{\partial \mathbb{E}_{r_0,\theta}[\overline{\mathcal{T}}]}{\partial t_{\rm{u}}}=w_{\rm{u}}\frac{\partial \mathbb{E}_{r_0,\theta}[\overline{R}_{\rm{u}}]}{\partial t_{\rm{u}}}+w_{\rm{d}}\frac{\partial \mathbb{E}_{r_0,\theta}[\overline{R}_{\rm{d}}]}{\partial t_{\rm{u}}}=0$. The expectation value of the average downlink throughput is $\mathbb{E}_{r_0,\theta}[\overline{R}_{\rm{d}}]=\iint_{r_0,\theta}\overline{R}_{\rm{d}}f_{\mathcal{R}_0}(r_0)f_{\Theta}(\theta){\rm{d}}\theta{\rm{d}}r_0$ and its derivation is equal to $\frac{\partial \mathbb{E}_{r_0,\theta}[\overline{R}_{\rm{d}}]}{\partial t_{\rm{u}}}=\frac{\partial }{\partial t_{\rm{u}}}\iint_{r_0,\theta}\overline{R}_{\rm{d}}f_{\mathcal{R}_0}(r_0)f_{\Theta}(\theta){\rm{d}}\theta{\rm{d}}r_0$. Since the function inside the integral is derivative on the range $(0,2r_{\rm{c}}/v)$, the derivation operator can go inside the integral as $\iint_{r_0,\theta}\frac{\partial \overline{R}_{\rm{d}}}{\partial t_{\rm{u}}}f_{\mathcal{R}_0}\!(r_0)f_{\Theta}(\theta){\rm{d}}\theta{\rm{d}}r_0$ \cite{hutton1984interchanging}, and this is the expectation value of the derivation of the average downlink throughput, i.e., $\mathbb{E}_{r_0,\theta}[\frac{\partial \overline{R}_{\rm{d}}}{\partial t_{\rm{u}}}]$. Thus, we can conclude that $\frac{\partial \mathbb{E}_{r_0,\theta}[\overline{R}_{\rm{d}}]}{\partial t_{\rm{u}}}\!=\!\mathbb{E}_{r_0,\theta}[\frac{\partial \overline{R}_{\rm{d}}}{\partial t_{\rm{u}}}]$. Using the same methodology for uplink throughput we have $\frac{\partial \mathbb{E}_{r_0,\theta}[\overline{R}_{\rm{u}}]}{\partial t_{\rm{u}}}=\mathbb{E}_{r_0,\theta}[\frac{\partial \overline{R}_{\rm{u}}}{\partial t_{\rm{u}}}]$. Then, the derivation of \eqref{equation17} can be expressed as:\vspace{-0.1cm}
\begin{equation}
	\label{equation31}
	\mathbb{E}_{r_0,\theta}\left[ \frac{\partial \overline{\mathcal{T}}}{\partial t_{\rm{u}}}\right] =w_{\rm{u}}\mathbb{E}_{r_0,\theta}\left[ \frac{\partial \overline{R}_{\rm{u}}}{\partial t_{\rm{u}}}\right] +w_{\rm{d}}\mathbb{E}_{r_0,\theta}\left[ \frac{\partial \overline{R}_{\rm{d}}}{\partial t_{\rm{u}}}\right] .
\end{equation}
Hence, the root of $\mathbb{E}_{r_0,\theta}[\frac{\partial \overline{\mathcal{T}}}{\partial t_{\rm{u}}}]=0$ will be the same as the root of $\frac{\partial \mathbb{E}_{r_0,\theta}[\overline{\mathcal{T}}]}{\partial t_{\rm{u}}}=0$.
	
Using the Leibniz integral rule the derivation of \eqref{equation18} can be obtained as:\vspace{-0.03cm}
\begin{equation}
	\label{equation32}
	\begin{aligned}
		&\dfrac{\partial {\overline{R}}_{\rm{u}}}{\partial t_{\rm{u}}}\!= \!\dfrac{-2(m+3)\widetilde{\mathcal{T}}_{\rm{u}}B_{{\rm{u}},n}\!}{Nt_{\rm{u}}^2}\!\left(\!\!1\!-\!\dfrac{\!2t_{\rm{fb}}}{t_{\rm{u}}}\! \right)\!\! \left(\dfrac{t_{\rm{u}}}{2(m+3)}\log_2\!\left(\!\dfrac{G_{\rm{u}}}{(r^2(t_{\rm{u}})+\!h^2)^{m+3}}\! \right)\!+\!\dfrac{r_0\cos\theta}{2v}\!\log_2 \left( r^2(t_{\rm{u}})\!+\!h^2\right)\right.\\
		&-\dfrac{\!(h^2\!+\!r_0^2\sin^2\!\theta)^{\frac{1}{2}}}{v\ln(2)}\!\tan^{-1}\!\!\left(\!\!\dfrac{vt_{\rm{u}}-r_0\cos\theta}{(h^2\!+\!r_0^2\sin^2\!\theta)^{\frac{1}{2}}} \!\!\right)\!\! -\dfrac{\!(h^2\!+\!r_0^2\sin^2\!\theta)^{\frac{1}{2}}}{v\ln(2)}\!\tan^{-1}\!\!\left(\!\!\dfrac{r_0\cos\theta}{(h^2\!+\!r_0^2\sin^2\!\theta)^{\frac{1}{2}}} \!\!\right)  \\
		& \left. -\dfrac{r_0\cos\theta}{2v}\!\log_2 \left( r_0^2\!+\!h^2\right)+\dfrac{t_{\rm{u}}}{\ln(2)}\right)  +\dfrac{\widetilde{\mathcal{T}}_{\rm{u}}B_{{\rm{u}},n}}{Nt_{\rm{u}}}\left(\!\!1\!-\!\dfrac{\!t_{\rm{fb}}}{t_{\rm{u}}}\! \right)\log_2\!\left(\!\dfrac{G_{\rm{u}}}{(r^2(t_{\rm{u}})+\!h^2)^{m+3}}\! \right) \\
		\end{aligned}
\end{equation}
Using the sum of inverse tangents formula, $\tan^{-1}(a)+\tan^{-1}(b)=\tan^{-1}\left( \frac{a+b}{1-ab} \right)$, \eqref{equation32} can be further simplified as:
\begin{equation}
	\label{equation33}
	\begin{aligned}
		&\dfrac{\partial {\overline{R}}_{\rm{u}}}{\partial t_{\rm{u}}}\!=\!\dfrac{-2(m+3)\widetilde{\mathcal{T}}_{\rm{u}}B_{{\rm{u}},n}}{Nt_{\rm{u}}^2}\!\left(\!\!1\!-\!\dfrac{\!2t_{\rm{fb}}}{t_{\rm{u}}}\! \right) \!\!\left(\!\dfrac{r_0\cos\theta}{2v}\!\log_2\!\! \left(\! \dfrac{r^2(t_{\rm{u}})\!+\!h^2}{r_0^2\!+\!h^2}\right)\right. \\ 
		&\resizebox{1\hsize}{!}{$\left.-\dfrac{\!(h^2\!+\!r_0^2\sin^2\!\theta)^{\frac{1}{2}}}{v\ln(2)}\!\tan^{-1}\!\!\left(\!\!\dfrac{\dfrac{vt_{\rm{u}}}{(h^2\!+\!r_0^2\sin^2\!\theta)^{\frac{1}{2}}}}{1\!-\dfrac{r_0\cos\theta(vt_{\rm{u}}\!-r_0\cos\theta)}{h^2+r_0^2\sin^2\!\theta}} \!\!\right)\!\!+\dfrac{t_{\rm{u}}}{\ln(2)}\!\!\right)\!\!+\dfrac{\widetilde{\mathcal{T}}_{\rm{u}}B_{{\rm{u}},n}t_{\rm{fb}}}{Nt_{\rm{u}}} \log_2\!\left(\!\dfrac{G_{\rm{u}}}{(r^2(t_{\rm{u}})\!+\!h^2)^{m+3}}\! \right).$}
		\end{aligned}
\end{equation}\vspace{-0.1cm}
This is the exact derivation of the average uplink achievable throughput respect to $t_{\rm{u}}$, however, for $vt_{\rm{u}}\ll h$, this equation can be further simplified. Substituting $r(t_{\rm{u}})=(r_0^2+v^2t_{\rm{u}}^2-2r_0vt_{\rm{u}}\cos\theta+h^2)^{1/2}$ in logarithm term, ignoring the small terms and using the approximation $\ln(1+x)\cong x$ for small values of $x$, we arrive  $\log_2\!\left(\!1\!+\!\frac{\!v^2t_{\rm{u}}^2-2r_0vt_{\rm{u}}\!\cos\!\theta}{r_0^2+h^2}\! \right)\!\cong\!\log_2\! \left(\!1\!-\!\frac{\!2r_0vt_{\rm{u}}\!\cos\!\theta}{r_0^2+h^2} \!\right)\!\cong\! \frac{-2r_0vt_{\rm{u}}\!\cos\!\theta}{\ln(2)\left( r_0^2+h^2\right) }$.
Considering the rule of small-angle approximation for inverse tangent, it can also be approximated by its first two terms of Taylor series as $\tan^{-1}(x)\cong x-x^3/3$ for small $x$. Noting that $t_{\rm{fb}}\ll t_{\rm{u}}$, the approximate derivation is given as follows:\vspace{-0.1cm}
\begin{equation}
	\label{equation34}
	\begin{aligned}
		&\dfrac{\partial {\overline{R}}_{\rm{u}}}{\partial t_{\rm{u}}}\!\cong
		\!\dfrac{-2(m+3)\widetilde{\mathcal{T}}_{\rm{u}}B_{{\rm{u}},n}}{\ln(2)Nt_{\rm{u}}^2}\!\left(\!1\!-\!\dfrac{\!2t_{\rm{fb}}}{t_{\rm{u}}}\! \right)\!\!\!\left(\!\dfrac{(vt_{\rm{u}})^3(h^2\!+r_0^2\sin^2\!\theta)^2}{3v(h^2+r_0^2)^3}\!+t_{\rm{u}} \right.\\
		&\left. -\dfrac{r_0^2\cos^2\!\theta t_{\rm{u}}}{r_0^2\!+\!h^2} -\!\dfrac{t_{\rm{u}}(h^2\!+\!r_0^2\sin^2\!\theta)}{h^2+r_0^2}\right)\!\!+\!\dfrac{\widetilde{\mathcal{T}}_{\rm{u}}B_{{\rm{u}},n}t_{\rm{fb}}}{Nt_{\rm{u}}^2}\! \log_2\!\left(\!\dfrac{G_{\rm{u}}}{\!(r_0^2\!+\!h^2)^{m+3}}\!\! \right)\\
		&=-\dfrac{\!2(m+3)\widetilde{\mathcal{T}}_{\rm{u}}B_{{\rm{u}},n}v^2(h^2+r_0^2\sin^2\!\theta)^2t_{\rm{u}}}{3N\ln(2)(h^2+r_0^2)^3}
		+\dfrac{\widetilde{\mathcal{T}}_{\rm{u}}B_{{\rm{u}},n}t_{\rm{fb}}}{Nt_{\rm{u}}^2} \! \log_2\!\left(\!\dfrac{G_{\rm{u}}}{\!(r_0^2+h^2)^{m+3}}\!\! \right)
	\end{aligned}
\end{equation}
	
Using the Leibniz integral rule to calculate the derivation of average downlink throughput, and the sum of inverse tangents formula to simplify it, the derivation of average downlink throughput is given as:\vspace{-0.2cm}
	\begin{equation}
		\label{equation37}
		\begin{aligned}
			&\dfrac{\partial \overline{R}_{\rm{d}}}{\partial t_{\rm{u}}}\!=\!\dfrac{\!\!-2(m+3)k_{\rm{req}}B_{{\rm{d}},n}}{\mathcal{K}t_{\rm{u}}^2}\!\! \left( \dfrac{\!r_0\!\cos\!\theta}{2v}\!\log_2 \left(\dfrac{r^2(t_{\rm{u}})+h^2}{r_0^2+h^2} \right)\!+\dfrac{t_{\rm{u}}}{\ln(2)} \right.\\
			&\left.  -\frac{(h^2\!+\!r_0^2\sin^2\!\theta)^{\frac{1}{2}}}{v\ln(2)}\tan^{-1}\!\!\left(\!\frac{\dfrac{vt_{\rm{u}}}{(h^2+r_0^2\sin^2\theta)^{\frac{1}{2}}}}{1\!-\!\dfrac{r_0\cos\theta(vt_{\rm{u}}\!-\!r_0\cos\theta)}{(h^2+r_0^2\sin^2\theta)}}\! \right) \right).
		\end{aligned}
	\end{equation}
This is the exact derivation of average downlink achievable throughput respect to $t_{\rm{u}}$, however, using the approximation rules for $vt_{\rm{u}}\ll h$,  the well-approximated derivation is given as follows:\vspace{-0.25cm}
\begin{equation}
	\label{equation38}
	\begin{aligned}
		&\frac{\partial \overline{R}_{\rm{d}}}{\partial t_{\rm{u}}}\cong \frac{-2(m+3)k_{\rm{req}}B_{{\rm{d}},n}}{\mathcal{K}t_{\rm{u}}^2} \left(-\frac{r_0^2\cos^2\theta t_{\rm{u}}}{\ln(2)(r_0^2+h^2)}-\frac{t_{\rm{u}}(h^2+r_0^2\sin^2\theta)}{\ln(2)(h^2+r_0^2)} \right.\\
		&\left. +\frac{(vt_{\rm{u}})^3(h^2+r_0^2\sin^2\theta)^2}{3v\ln(2)(h^2+r_0^2)^3}+\frac{t_{\rm{u}}}{\ln(2)} \right)=\frac{-2(m+3)k_{\rm{req}}B_{{\rm{d}},n}v^2t_{\rm{u}}(h^2+r_0^2\sin^2\theta)^2}{3\mathcal{K}\ln(2)(h^2+r_0^2)^3}.
		\end{aligned}
\end{equation}
	
The exact optimum time, $t_{\rm{u,opt}}$, can be obtained numerically by solving \eqref{equation21} after substituting $\frac{\partial \overline{R}_{\rm{d}}}{\partial t_{\rm{u}}}$ and $\frac{\partial \overline{R}_{\rm{u}}}{\partial t_{\rm{u}}}$ given in \eqref{equation32} and \eqref{equation37}. However, we can approximately obtain a closed form for optimum update interval denoted as $\widetilde{t}_{\rm{u,opt}}$ by using \eqref{equation34} and  \eqref{equation38}. Taking into account that $vt_{\rm{u}}\ll h$ the closed solution form for optimum update interval is given as:
\vspace{-5pt}
\begin{equation*}
	\label{equation39}
	\begin{aligned}
		\!\!\!\widetilde{t}_{\rm{u,opt}}\!\cong\!\! \left(\!\dfrac{\frac{3\!\ln(2)}{2(m+3)}w_{\rm{u}}t_{\rm{fb}}\widetilde{\mathcal{T}}_{\rm{u}}B_{{\rm{u}},n}C_1 }{w_{\rm{d}}v^2NR_{\rm{req}}+C_2w_{\rm{u}}v^2\widetilde{\mathcal{T}}_{\rm{u}}B_{{\rm{u}},n}}\! \right)^{\!\!\frac{1}{3}},
	\end{aligned}
\end{equation*}
where
\begin{equation*}
	\label{equation40}
	\begin{aligned}
	&C_1=\dfrac{\mathbb{E}_{r_0}\!\!\left[\log_2\!\!\left(\!\dfrac{G_{\rm{u}}}{(r_0^2+h^2)^{m+3}}\!\right)\right]\mathbb{E}_{r_0}\!\!\left[ \log_2\!\!\left(\!\dfrac{G}{(r_0^2+h^2)^{m+3}}\!\right)\!\right]}{\mathbb{E}_{r_0,\theta}\!\!\left[\dfrac{(h^2+r_0^2\sin^2\theta)^2}{(h^2+r_0^2)^3}\right] }, \ 
	C_2=\mathbb{E}_{r_0}\!\!\left[ \log_2\!\!\left(\!\dfrac{G}{(r_0^2+h^2)^{m+3}}\!\right)\!\right].
	\end{aligned}
\end{equation*}

\footnotesize
\bibliographystyle{IEEEtran}
\bibliography{IEEEabrv.bib,Ref.bib}

% Generated by IEEEtranTCOM.bst, version: 1.13 (2008/09/30)
\begin{thebibliography}{10}
\baselineskip 12pt
\providecommand{\url}[1]{#1}
\csname url@samestyle\endcsname
\providecommand{\newblock}{\relax}
\providecommand{\bibinfo}[2]{#2}
\providecommand{\BIBentrySTDinterwordspacing}{\spaceskip=0pt\relax}
\providecommand{\BIBentryALTinterwordstretchfactor}{4}
\providecommand{\BIBentryALTinterwordspacing}{\spaceskip=\fontdimen2\font plus
\BIBentryALTinterwordstretchfactor\fontdimen3\font minus
  \fontdimen4\font\relax}
\providecommand{\BIBforeignlanguage}[2]{{%
\expandafter\ifx\csname l@#1\endcsname\relax
\typeout{** WARNING: IEEEtran.bst: No hyphenation pattern has been}%
\typeout{** loaded for the language `#1'. Using the pattern for}%
\typeout{** the default language instead.}%
\else
\language=\csname l@#1\endcsname
\fi
#2}}
\providecommand{\BIBdecl}{\relax}
\BIBdecl

\bibitem{Cisco}
Cisco, ``{Cisco VNI Mobile Forecast (2015 -- 2020)},'' \emph{white paper at
  Cisco.com}, Feb. 2016.

\bibitem{Haas}
H.~Haas, L.~Yin, Y.~Wang, and C.~Chen, ``{What is LiFi?}'' \emph{J. Lightwave
  Technol.}, vol.~34, no.~6, pp. 1533--1544, Mar. 2016.

\bibitem{WuVLC5G}
S.~Wu, H.~Wang, and C.~H. Youn, ``{Visible Light Communications for 5G Wireless
  Networking Systems: From Fixed to Mobile Communications},'' \emph{{IEEE}
  Network}, vol.~28, no.~6, pp. 41--45, Nov. 2014.

\bibitem{LFchen2014performance}
X.~Chen \emph{et~al.}, ``{Performance Analysis and Optimization for
  Interference Alignment Over MIMO Interference Channels With Limited
  Feedback},'' \emph{IEEE Trans. Signal Process.}, vol.~62, no.~7, pp.
  1785--1795, 2014.

\bibitem{LFHeath1}
R.~Bhagavatula and R.~W. Heath, ``{Adaptive Limited Feedback For Sum-rate
  Maximizing Beamforming in Cooperative Multicell Systems},'' \emph{IEEE Trans.
  Signal Process.}, vol.~59, no.~2, pp. 800--811, 2011.

\bibitem{LFHeath2}
D.~J. Ryan, I.~B. Collings, I.~V.~L. Clarkson, and R.~W. Heath, ``{Performance
  of Vector Perturbation Multiuser MIMO Systems With Limited Feedback},''
  \emph{IEEE Trans. on Commun.}, vol.~57, no.~9, 2009.

\bibitem{love2008overview}
D.~J. Love, R.~W. Heath, V.~K. Lau, D.~Gesbert, B.~D. Rao, and M.~Andrews,
  ``{An Overview of Limited Feedback in Wireless Communication Systems},''
  \emph{IEEE J. Sel. Areas Commun.}, vol.~26, no.~8, pp. 1341--1365, 2008.

\bibitem{jang2006throughput}
E.~W. Jang, Y.~Cho, and J.~M. Cioffi, ``{SPC12-4: Throughput Optimization for
  Continuous Flat Fading MIMO Channels with Estimation Error},'' in \emph{Proc.
  2006 IEEE Globecom Conf.}, pp. 1--5.

\bibitem{pattanayak2016sinr}
P.~Pattanayak and P.~Kumar, ``{SINR Based Limited Feedback Scheduling For
  MIMO-OFDM Heterogeneous Broadcast Networks},'' in \emph{Proc. 2016 IEEE
  Twenty Second National Conf. Commun. (NCC)}, 2016, pp. 1--6.

\bibitem{mokari2016limited}
N.~Mokari, F.~Alavi, S.~Parsaeefard, and T.~Le-Ngoc, ``{Limited-Feedback
  Resource Allocation in Heterogeneous Cellular Networks},'' \emph{IEEE Trans.
  Veh. Technol.}, vol.~65, no.~4, pp. 2509--2521, 2016.

\bibitem{leinonen2008performance}
J.~Leinonen, J.~Hamalainen, and M.~Juntti, ``{Performance Analysis of Downlink
  OFDMA Frequency Scheduling With Limited Feedback},'' in \emph{Proc. 2008 IEEE
  Int. Conf. Commun. (ICC)}, pp. 3318--3322.

\bibitem{vu2007capacity}
M.~Vu and A.~Paulraj, ``{On The Capacity of MIMO Wireless Channels With Dynamic
  CSIT},'' \emph{IEEE J. Sel. Areas Commun.}, vol.~25, no.~7, pp. 1269--1283,
  2007.

\bibitem{sun2003asymptotic}
Y.~Sun and M.~Honig, ``{Asymptotic Capacity of Multicarrier Transmission Over a
  Fading Channel With Feedback},'' in \emph{Proc. 2003 IEEE Int. Symp. Inf.
  Theory}, p.~40.

\bibitem{sun2003minimum}
------, ``{Minimum Feedback Rates For Multicarrier Transmission With Correlated
  Frequency-selective Fading},'' in \emph{Proc. 2003 IEEE Globecom Conf.},
  vol.~3, pp. 1628--1632.

\bibitem{agarwal2008multi}
M.~Agarwal, D.~Guo, and M.~L. Honig, ``{Multi-carrier Transmission With Limited
  Feedback: Power Loading Over Sub-channel Groups},'' in \emph{Proc. 2008 IEEE
  Int. Conf. Commun. (ICC)}, 2008, pp. 981--985.

\bibitem{svedman2004simplified}
P.~Svedman, S.~K. Wilson, L.~J. Cimini, and B.~Ottersten, ``{A Simplified
  Opportunistic Feedback and Scheduling Scheme for OFDM},'' in \emph{Proc. 2004
  IEEE 59th Veh. Technol. Conf. (VTC)}, vol.~4, pp. 1878--1882.

\bibitem{SaraScheduling}
S.~K. Wilson and J.~Holliday, ``{Scheduling Methods for Multi-user Optical
  Wireless Asymmetrically-clipped OFDM},'' \emph{J. Commun. and Netw.},
  vol.~13, no.~6, pp. 655--663, Dec 2011.

\bibitem{floren2003effect}
F.~Flor{\'e}n, O.~Edfors, and B.-A. Molin, ``{The Effect of Feedback
  Quantization on the Throughput of a Multiuser Diversity Scheme},'' in
  \emph{Proc. 2003 IEEE Globecom Conf.}, vol.~1, pp. 497--501.

\bibitem{gesbert2003selective}
D.~Gesbert and M.-S. Alouini, ``{Selective Multi-user Diversity},'' in
  \emph{Proc. 2003 3rd IEEE Int. Symp. Signal Proc. and Inf. Technol.
  (ISSPIT)}, pp. 162--165.

\bibitem{gesbert2004much}
------, ``{How Much Feedback Is Multi-user Diversity Really Worth?}'' in
  \emph{Proc. 2004 IEEE Int. Conf. Commun. (ICC)}, vol.~1, pp. 234--238.

\bibitem{sanayei2007opportunistic}
S.~Sanayei and A.~Nosratinia, ``{Opportunistic Downlink Transmission With
  Limited Feedback},'' \emph{IEEE Trans. Inf. Theory}, vol.~53, no.~11, pp.
  4363--4372, 2007.

\bibitem{chen2006large}
J.~Chen, R.~A. Berry, and M.~L. Honig, ``{Large System Performance of Downlink
  OFDMA With Limited Feedback},'' in \emph{Proc. 2006 IEEE Int. Symp. Inf.
  Theory}, pp. 1399--1403.

\bibitem{hassel2007threshold}
V.~Hassel, D.~Gesbert, M.-S. Alouini, and G.~E. Oien, ``{A Threshold-based
  Channel State Feedback Algorithm For Modern Cellular Systems},'' \emph{IEEE
  Trans. Wireless Commun.}, vol.~6, no.~7, p. 2422, 2007.

\bibitem{HartmanPatent}
P.~A. Hartman, ``{Cellular Frequency Reuse Cell Plan},'' U.S. Patent 5, 247,
  699, Sep. 21, 1993.

\bibitem{chen2015fractional}
C.~Chen, S.~Videv, D.~Tsonev, and H.~Haas, ``{Fractional Frequency Reuse in
  DCO-OFDM-based Optical Attocell Networks},'' \emph{J. Lightw. Technol.},
  vol.~33, no.~19, pp. 3986--4000, 2015.

\bibitem{Dinc}
E.~Dinc, O.~Ergul, and O.~B. Akan, ``{Soft Handover in OFDMA Based Visible
  Light Communication Networks},'' in \emph{Proc. 2015 IEEE 82th Veh. Technol.
  Conf. (VTC)}, pp. 1--5.

\bibitem{SoftHandoverChoi}
H.~H. Choi, ``{An Optimal Handover Decision for Throughput Enhancement},''
  \emph{IEEE Commun. Lett.}, vol.~14, no.~9, pp. 851--853, 2010.

\bibitem{ChenCheng}
C.~Chen, D.~A.~Basnayaka, and H.~Haas, ``{Downlink Performance of Optical
  Attocell Networks},'' \emph{J. Lightw. Technol.}, vol.~34, no.~1, pp.
  137--156, Jan. 2016.

\bibitem{barry1993simulation}
J.~R. Barry, J.~M. Kahn, W.~J. Krause, E.~A. Lee, and D.~G. Messerschmitt,
  ``{Simulation of Multipath Impulse Response for Indoor Wireless Optical
  Channels},'' \emph{IEEE J. Sel. Areas Commun}, vol.~11, no.~3, pp. 367--379,
  1993.

\bibitem{le2009100}
H.~Le~Minh \emph{et~al.}, ``{100-Mb/s NRZ Visible Light Communications Using a
  Postequalized White LED},'' \emph{IEEE Photon. Technol. Lett.}, vol.~21,
  no.~15, pp. 1063--1065, 2009.

\bibitem{Bettstetter}
C.~Bettstetter, H.~Hartenstein, and X.~Perez-Costa, ``{Stochastic Properties of
  the Random Waypoint Mobility Model},'' \emph{ACM Wireless Netw.}, vol.~10,
  no.~5, pp. 555--567, Sep. 2004.

\bibitem{Andrews}
X.~Lin \emph{et~al.}, ``{Towards Understanding the Fundamentals of Mobility in
  Cellular Networks},'' \emph{IEEE Trans. Wireless Commun.}, vol.~12, no.~4,
  pp. 1686--1698, April 2013.

\bibitem{EsaRWP}
E.~Hyyti{\"a} and J.~Virtamo, ``{Random Waypoint Mobility Model in Cellular
  Networks},'' \emph{Wireless Networks}, vol.~13, no.~2, pp. 177--188, 2007.

\bibitem{armstrong2008comparison}
J.~Armstrong and B.~J. Schmidt, ``{Comparison of Asymmetrically Clipped Optical
  OFDM and DC-biased Optical OFDM in AWGN},'' \emph{IEEE Commun. Lett.},
  vol.~12, no.~5, pp. 343--345, 2008.

\bibitem{dissanayake2013comparison}
S.~D. Dissanayake and J.~Armstrong, ``{Comparison of ACO-OFDM, DCO-OFDM and
  ADO-OFDM in IM/DD Systems},'' \emph{J. Lightw. Technol.}, vol.~31, no.~7, pp.
  1063--1072, 2013.

\bibitem{jalali2000data}
A.~Jalali, R.~Padovani, and R.~Pankaj, ``{Data Throughput of CDMA-HDR a High
  Efficiency-high Data Rate Personal Communication Wireless System},'' in
  \emph{Proc. 2000 IEEE 51th Veh. Technol. Conf. (VTC)}, vol.~3, pp.
  1854--1858.

\bibitem{kim2002proportionally}
K.~Kim, H.~Kim, and Y.~Han, ``{A Proportionally Fair Scheduling Algorithm With
  QoS and Priority in 1xEV-DO},'' in \emph{Proc. 2002 13th IEEE Int. Symp. on
  Personal Indoor and Mobile Radio Commun (PIMRC)}, vol.~5, pp. 2239--2243.

\bibitem{dimitrov2013information}
S.~Dimitrov and H.~Haas, ``{Information Rate of OFDM-Based Optical Wireless
  Communication Systems With Nonlinear Distortion},'' \emph{Journal of
  Lightwave Technology}, vol.~31, no.~6, pp. 918--929, 2013.

\bibitem{IEEEMACStand}
\emph{{IEEE 802.11: Wireless LAN Medium Access Control (MAC) and Physical Layer
  (PHY) Specifications}}, IEEE-SA Std., Rev. 2012.

\bibitem{bianchi2000performance}
G.~Bianchi, ``{Performance Analysis of the IEEE 802.11 Distributed Coordination
  Function},'' \emph{IEEE J. Sel. Areas Commun.}, vol.~18, no.~3, pp. 535--547,
  2000.

\bibitem{sanayei2005exploiting}
S.~Sanayei and A.~Nosratinia, ``{Exploiting Multiuser Diversity With Only 1-bit
  Feedback},'' in \emph{Proc. 2005 IEEE Wireless Commun. Netw. Conf.}, vol.~2,
  pp. 978--983.

\bibitem{DehghaniSoltani2015}
M.~Dehghani~Soltani, X.~Wu, M.~Safari, and H.~Haas, ``{On Limited Feedback
  Resource Allocation for Visible Light Communication Networks},'' in
  \emph{Proc. 2015 ACM 2nd Int. Workshop on Visible Light Communications
  Systems (VLCS)}, pp. 27--32.

\bibitem{MDSPIMRC2017}
M.~Dehghani~Soltani, M.~Safari, and H.~Haas, ``{On Throughput Maximization
  Based on Optimal Update Interval in LiFi Networks},'' \emph{Accepted in Proc.
  2017 28th IEEE Int. Symp. on Personal Indoor and Mobile Radio Commun
  (PIMRC)}.

\bibitem{cirik2015weighted}
A.~C. Cirik, R.~Wang, Y.~Hua, and M.~Latva-aho, ``{Weighted Sum-rate
  Maximization For Full-duplex MIMO Interference Channels},'' \emph{IEEE Trans.
  Commun.}, vol.~63, no.~3, pp. 801--815, 2015.

\bibitem{aquilina2017weighted}
P.~Aquilina, A.~Cirik, and T.~Ratnarajah, ``{Weighted Sum Rate Maximization in
  Full-Duplex Multi-User Multi-Cell MIMO Networks},'' \emph{IEEE Trans.
  Commun.}, 2017.

\bibitem{bohannon1997comfortable}
R.~W. Bohannon, ``{Comfortable and Maximum Walking Speed of Adults Aged 20--79
  Years: Reference Values and Determinants},'' \emph{Age and ageing}, vol.~26,
  no.~1, pp. 15--19, 1997.

\bibitem{3gpp.36.213}
``\textit{Physical Layer Procedures},'' {3GPP}, TS {36.213}, 2016.

\bibitem{hutton1984interchanging}
J.~E. Hutton and P.~I. Nelson, ``{Interchanging the Order of Differentiation
  and Stochastic Integration},'' \emph{Stochastic processes and their
  applications}, vol.~18, no.~2, pp. 371--377, 1984.

\end{thebibliography}

\end{document}